\documentclass[notitlepage,a4paper,aps,prd,onecolumn,superscriptaddress,nofootinbib,groupedaddress,longbibliography]{revtex4-2}
\usepackage{amsmath}
\usepackage{amsfonts}
\usepackage{amssymb}
\usepackage{accents}
\usepackage[utf8]{inputenc}
\usepackage[T1]{fontenc}
\usepackage[colorlinks=true]{hyperref}
\allowdisplaybreaks

\newcommand{\lc}[1]{\accentset{\circ}{#1}}
\newcommand{\st}[1]{\underaccent{Q}{#1}}
\newcommand{\mt}[1]{\underaccent{T}{#1}}
\newcommand{\dd}{\mathrm{d}}
\newcommand{\jf}[1]{\accentset{\mathfrak{J}}{#1}}
\newcommand{\ef}[1]{\accentset{\mathfrak{E}}{#1}}

\begin{document}

\title{Field transformations and invariant quantities in scalar-teleparallel theories of gravity}

\author{Manuel Hohmann}
\email{manuel.hohmann@ut.ee}
\affiliation{Laboratory of Theoretical Physics, Institute of Physics, University of Tartu, W. Ostwaldi 1, 50411 Tartu, Estonia}

\begin{abstract}
We study transformations of the dynamical fields - a metric, a flat affine connection and a scalar field - in scalar-teleparallel gravity theories. The theories we study belong either to the general teleparallel setting, where no further condition besides vanishing curvature is imposed on the affine connection, or the symmetric or metric teleparallel gravity, where one also imposes vanishing torsion or nonmetricity, respectively. For each of these three settings, we find a general class of scalar-teleparallel action functionals which retain their form under the aforementioned field transformations. This is achieved by generalizing the constraint of vanishing torsion or nonmetricity to non-vanishing, but algebraically constrained torsion or nonmetricity. We find a number of invariant quantities which characterize these theories independently of the choice of field variables, and relate these invariants to analogues of the conformal frames known from scalar-curvature gravity. Using these invariants, we are able to identify a number of physically relevant subclasses of scalar-teleparallel theories. We also generalize our results to multiple scalar fields, and speculate on further extended theories with non-vanishing, but algebraically constrained curvature.
\end{abstract}

\maketitle

%\tableofcontents

\section{Introduction}\label{sec:intro}
Our current understanding of the gravitational interaction is largely based on general relativity, which describes the gravitational interaction through the curvature of the Levi-Civita connection of a Lorentzian metric. It is a highly successful theory, whose predictions agree with observations on different scales, including the solar system, black holes and gravitational waves~\cite{Baker:2014zba}. However successful, it leaves a number of open questions to be answered: its application to cosmology, which leads to the so-called $\Lambda$CDM model describing 95\% of the matter-energy content of the universe as dark energy in form of a cosmological constant \(\Lambda\) and cold dark matter (CDM), is challenged by observational tensions~\cite{Planck:2018vyg,DiValentino:2021izs}, and any attempts to conclusively quantize general relativity or obtain a unified theory of gravity and the other fundamental interactions have so far remained unsuccessful. Modified theories of gravity, which aim to solve these open problems unanswered by general relativity, are therefore an actively studied topic in modern physics~\cite{Nojiri:2006ri,Nojiri:2010wj,Faraoni:2010pgm,Clifton:2011jh,Nojiri:2017ncd,Bull:2015stt,Heisenberg:2018vsk,CANTATA:2021ktz,Odintsov:2023weg,Pfeifer:2023cgd}.

A large class of theories which is subject to current research are teleparallel gravity theories~\cite{Hohmann:2022mlc}. These theories fall into the more general class of metric-affine gravity theories~\cite{Hehl:1994ue}, as they are based on the assumption that another fundamental field variable besides the metric is an independent affine connection, which is imposed to be flat. There are three types of teleparallel theories, conventionally called general, symmetric and metric teleparallel gravity, where the latter two are obtained from the general class by imposing additional constraints on the flat affine connection - either vanishing torsion or vanishing nonmetricity. The remaining, non-vanishing tensorial properties of the connection, i.e., nonmetricity in the symmetric case, torsion in the metric case and both of them in the general case, then carry the gravitational interaction, and therefore take the role which is taken by the curvature in general relativity. Through a suitable choice of the action functional, each of these three types of teleparallel theories allows the construction of a teleparallel equivalent of general relativity, where equivalence is to be understood as leading to identical field equations for the metric and thus also identical solutions~\cite{Maluf:2013gaa,Nester:1998mp,BeltranJimenez:2019odq,Boehmer:2021aji,Capozziello:2022zzh}.

One of the reasons for studying teleparallel gravity theories is the fact that they admit modifications of their general relativity equivalents whose dynamics differs from similar modifications of general relativity, leading to new possibilities to address the aforementioned observational tensions~\cite{Cai:2015emx,Bahamonde:2021gfp,Heisenberg:2023lru}. One of the most simple modifications, which is motivated by a similar class of modifications of general relativity~\cite{Damour:1992we,Faraoni:2004pi,Fujii:2003pa}, is the addition of one or more scalar fields, which leads to the notion of scalar-torsion~\cite{Linder:2010py,Geng:2011aj,Hohmann:2018rwf,Hohmann:2018vle,Hohmann:2018dqh,Hohmann:2018ijr}, scalar-nonmetricity~\cite{Jarv:2018bgs,Runkla:2018xrv} and general scalar-teleparallel~\cite{Hohmann:2022mlc,Heisenberg:2022mbo} theories of gravity. Among these theories one finds large classes which are able to address open questions in cosmology, while being consistent with observations in the solar system, described by their post-Newtonian limit.

Numerous aspects of scalar-teleparallel gravity theories are still to be studied and thoroughly understood. One such aspect is the conformal frame freedom, which is known from scalar-curvature gravity theories~\cite{Flanagan:2004bz}, and has become an actively debated topic~\cite{Catena:2006bd,Faraoni:2006fx,Deruelle:2010ht,Chiba:2013mha,Postma:2014vaa,Faraoni:1998qx,Capozziello:2010sc,Rondeau:2017xck,Wetterich:2019qzx,Bamber:2022eoy,Chakraborty:2023kel}. An interesting consequence of this freedom is the existence of a number of invariant quantities~\cite{Jarv:2014hma,Kuusk:2015dda}, which allow the description of phenomenological properties of scalar-curvature theories independently of the choice of the conformal frame~\cite{Jarv:2015kga,Kuusk:2016rso,Jarv:2016sow,Karam:2017zno}. It has been shown that analogously constructed scalar-torsion~\cite{Hohmann:2018ijr} and scalar-nonmetricity~\cite{Jarv:2018bgs,Runkla:2018xrv} theories admit a similar freedom of conformal transformations, either acting on the tetrad defining the metric and the connection or the metric alone, thus maintaining the constraints of vanishing curvature and either vanishing nonmetricity or vanishing torsion. Another open question in modified teleparallel gravity theories is the appearance of a strong coupling problem~\cite{Golovnev:2018wbh,Golovnev:2020zpv,BeltranJimenez:2020fvy,Blagojevic:2020dyq,Guzman:2019oth,BeltranJimenez:2019nns,Bahamonde:2022ohm,Golovnev:2020nln,Li:2023fto}, which manifests itself by the absence of dynamical degrees of freedom in the linear perturbation theory around highly symmetric backgrounds, which reappear in higher order perturbations, thus challenging the validity of the perturbative approach to solving the field equations.

The aim of this article is to address some of the aforementioned open questions in scalar-teleparallel gravity. Our main focus is on the conformal frame freedom, which we aim to study in all three types of scalar-teleparallel theories. For this purpose, we need to generalize and unify the different notions of conformal transformations in scalar-torsion and scalar-nonmetricity theories - either transforming the connection to maintain vanishing nonmetricity or leaving the connection unchanged to maintain vanishing torsion - and extend their application to the general scalar-teleparallel class of theories. Transformations of this type fall into a more general class of transformations in metric-affine gravity theories~\cite{Iosifidis:2018zwo}. We will see that these extended transformations lead to a natural generalization of scalar-torsion and scalar-nonmetricity theories to also include non-vanishing nonmetricity and torsion, respectively. The latter may serve as an additional starting point to also address the question of strong coupling, as it leads to more general background geometries around which perturbations may be studied. In particular, we study the existence of invariant quantities similar to the scalar-curvature case and the possibility to use these for a frame-independent characterization of scalar-teleparallel gravity theories. We consider a single scalar field at first, and then generalize our results to multiple scalar fields, where we also discuss the difficulties arising from this generalization.

The outline of this article is as follows. In section~\ref{sec:geom}, we briefly review the dynamical fields in scalar-teleparallel gravity theories, and introduce the class of field transformations we study in this article. We then provide a brief overview of the different types of scalar-teleparallel gravity actions in section~\ref{sec:action}, where we also define the matter coupling we consider here. The transformation of these actions is studied in section~\ref{sec:actiontrans}. We make use of these transformations in section~\ref{sec:invariant}, where we identify a number of invariant quantities. These are further used in section~\ref{sec:frames} to state the aforementioned action functionals in particular frames. We generalize our results to multiple scalar fields in section~\ref{sec:multi}, where we also point out necessary restrictions arising for this generalization. As an application of our findings, we provide an invariant characterization of several subclasses of (multi-)scalar-teleparallel theories in section~\ref{sec:char}. We end with a conclusion in section~\ref{sec:conclusion}.

\section{Field transformations in scalar-teleparallel geometry}\label{sec:geom}
We will start our discussion with a brief review of the scalar-teleparallel geometry. Throughout this article, we will assume that the fundamental fields mediating the gravitational interaction are given by a metric \(g_{\mu\nu}\), an affine connection with coefficients \(\Gamma^{\mu}{}_{\nu\rho}\), which is imposed to be flat,
\begin{equation}\label{eq:curvature}
R^{\mu}{}_{\nu\rho\sigma} = \partial_{\rho}\Gamma^{\mu}{}_{\nu\sigma} - \partial_{\sigma}\Gamma^{\mu}{}_{\nu\rho} + \Gamma^{\mu}{}_{\tau\rho}\Gamma^{\tau}{}_{\nu\sigma} - \Gamma^{\mu}{}_{\tau\sigma}\Gamma^{\tau}{}_{\nu\rho} \equiv 0\,,
\end{equation}
as well as a scalar field \(\phi\). In presence of a metric, the connection is fully characterized by the torsion
\begin{equation}\label{eq:torsion}
T^{\mu}{}_{\nu\rho} = \Gamma^{\mu}{}_{\rho\nu} - \Gamma^{\mu}{}_{\nu\rho}\,,
\end{equation}
as well as the nonmetricity
\begin{equation}\label{eq:nonmetricity}
Q_{\mu\nu\rho} = \nabla_{\mu}g_{\nu\rho} = \partial_{\mu}g_{\nu\rho} - \Gamma^{\sigma}{}_{\nu\mu}g_{\sigma\rho} - \Gamma^{\sigma}{}_{\rho\mu}g_{\nu\sigma}\,.
\end{equation}
This can be seen by defining the contortion
\begin{equation}\label{eq:contortion}
K^{\mu}{}_{\nu\rho} = \frac{1}{2}\left(T_{\nu}{}^{\mu}{}_{\rho} + T_{\rho}{}^{\mu}{}_{\nu} - T^{\mu}{}_{\nu\rho}\right)\,,
\end{equation}
and the disformation
\begin{equation}\label{eq:disformation}
L^{\mu}{}_{\nu\rho} = \frac{1}{2}\left(Q^{\mu}{}_{\nu\rho} - Q_{\nu}{}^{\mu}{}_{\rho} - Q_{\rho}{}^{\mu}{}_{\nu}\right)\,.
\end{equation}
With the help of these two quantities, it is possible to write the difference between the coefficients of the teleparallel and Levi-Civita connections, i.e., the Christoffel symbols
\begin{equation}\label{eq:levicivita}
\lc{\Gamma}^{\mu}{}_{\nu\rho} = \frac{1}{2}g^{\mu\sigma}\left(\partial_{\nu}g_{\sigma\rho} + \partial_{\rho}g_{\nu\sigma} - \partial_{\sigma}g_{\nu\rho}\right)\,,
\end{equation}
as
\begin{equation}\label{eq:conndec}
\Gamma^{\mu}{}_{\nu\rho} - \lc{\Gamma}^{\mu}{}_{\nu\rho} = M^{\mu}{}_{\nu\rho} = K^{\mu}{}_{\nu\rho} + L^{\mu}{}_{\nu\rho}\,.
\end{equation}
Here, \(M^{\mu}{}_{\nu\rho}\) is called the distortion. Note that we denote quantities which are defined through the Levi-Civita connection with a circle on top in order to distinguish them from the corresponding quantities defined through the teleparallel connection.

In the following, we will study transformations of the aforementioned fundamental fields. For the scalar field \(\phi\), we consider a general redefinition of the form
\begin{equation}\label{eq:scaltrans}
\bar{\phi} = f(\phi)
\end{equation}
with an arbitrary, invertible function \(f\). For the metric, we consider a conformal transformation given by
\begin{equation}\label{eq:mettrans}
\bar{g}_{\mu\nu} = g_{\mu\nu}e^{2\gamma(\phi)}\,,
\end{equation}
where \(\gamma\) is an arbitrary function. Finally, for the connection we study transformations of the form
\begin{equation}\label{eq:conntrans}
\bar{\Gamma}^{\mu}{}_{\nu\rho} = \Gamma^{\mu}{}_{\nu\rho} + \zeta(\phi)\delta^{\mu}_{\nu}\phi_{,\rho}\,,
\end{equation}
with another arbitrary function \(\zeta\) of the scalar field, for reasons which shall become as follows. First, note that it follows from the flatness of the connection that it has a path-independent parallel transport on simply connected regions, from which follows that locally its coefficients can be written as
\begin{equation}
\Gamma^{\mu}{}_{\nu\rho} = (\Lambda^{-1})^{\mu}{}_{\sigma}\partial_{\rho}\Lambda^{\sigma}{}_{\nu}\,.
\end{equation}
This parallel transport changes by a scalar-field dependent factor \(\xi(\phi)\) if we replace the coefficients \(\Lambda^{\mu}{}_{\nu}\) by
\begin{equation}
\bar{\Lambda}^{\mu}{}_{\nu} = \xi(\phi)\Lambda^{\mu}{}_{\nu}\,,
\end{equation}
from which follows that the connection defined by this new parallel transport has the coefficients
\begin{equation}
\bar{\Gamma}^{\mu}{}_{\nu\rho} = (\bar{\Lambda}^{-1})^{\mu}{}_{\sigma}\partial_{\rho}\bar{\Lambda}^{\sigma}{}_{\nu} = \Gamma^{\mu}{}_{\nu\rho} + \xi'(\phi)\delta^{\mu}_{\nu}\phi_{,\rho}\,,
\end{equation}
which becomes equal to the transformation~\ref{eq:conntrans} with \(\zeta(\phi) = \xi'(\phi)\), reflecting the fact that a constant factor \(\xi\), which does not depend on \(\phi\), does not change the connection. It follows that this transformation retains the flatness of the connection: by direct calculation, one finds
\begin{equation}\label{eq:curtrans}
\bar{R}^{\mu}{}_{\nu\rho\sigma} = \partial_{\rho}\bar{\Gamma}^{\mu}{}_{\nu\sigma} - \partial_{\sigma}\bar{\Gamma}^{\mu}{}_{\nu\rho} + \bar{\Gamma}^{\mu}{}_{\tau\rho}\bar{\Gamma}^{\tau}{}_{\nu\sigma} - \bar{\Gamma}^{\mu}{}_{\tau\sigma}\bar{\Gamma}^{\tau}{}_{\nu\rho} = R^{\mu}{}_{\nu\rho\sigma} \equiv 0\,.
\end{equation}
Further, we find the transformation of the torsion
\begin{equation}\label{eq:tortrans}
\bar{T}^{\mu}{}_{\nu\rho} = \bar{\Gamma}^{\mu}{}_{\rho\nu} - \bar{\Gamma}^{\mu}{}_{\nu\rho} = T^{\mu}{}_{\nu\rho} - 2\zeta(\phi)\delta^{\mu}_{[\nu}\phi_{,\rho]}
\end{equation}
and the nonmetricity
\begin{equation}\label{eq:nomtrans}
\bar{Q}_{\mu\nu\rho} = \bar{\nabla}_{\mu}\bar{g}_{\nu\rho} = e^{2\gamma(\phi)}[Q_{\mu\nu\rho} - 2(\zeta(\phi) - \gamma'(\phi))g_{\nu\rho}\phi_{,\mu}]\,.
\end{equation}
Hence, we see that if the initial teleparallel connection is torsion-free, \(T^{\mu}{}_{\nu\rho} = 0\), then this property is also retained by the transformed connection, provided that we choose \(\zeta(\phi) \equiv 0\). Similarly, if we start from a metric-compatible connection, \(Q_{\mu\nu\rho} = 0\), then this holds also for the transformed connection, if we choose \(\zeta(\phi) \equiv \gamma'(\phi)\). We will make use of these choices when we consider the transformation of metric and symmetric teleparallel geometries and gravity theories in the following sections. Further, we will use the convention that indices of transformed (barred) tensor fields are raised and lowered with the transformed metric \(\bar{g}_{\mu\nu}\), while indices of original (unbarred) tensor fields are raised and lowered with the original metric \(g_{\mu\nu}\).

\section{Action}\label{sec:action}
We will now introduce a number of scalar-teleparallel gravitational theories, whose behavior under the previously introduced field transformations we will study in the remainder of this article. We start by introducing the general form of the matter action, which defines the coupling between the matter and gravitational fields, in section~\ref{ssec:mataction}. For the gravitational part of the action, we have to distinguish three different cases, depending on whether we impose vanishing torsion or nonmetricity, or allow for both of them to be non-vanishing. We start with the latter case, known as general teleparallel gravity, in section~\ref{ssec:genaction}. We then impose vanishing torsion in section~\ref{ssec:symaction} and vanishing nonmetricity in section~\ref{ssec:metaction}.

\subsection{Matter part}\label{ssec:mataction}
We start our discussion of scalar-teleparallel gravity theories by providing a general matter action, which we choose to be of the form
\begin{equation}\label{eq:mataction}
S_{\text{m}}\left[g_{\mu\nu}, \Gamma^{\mu}{}_{\nu\rho}, \phi, \chi^I\right] = \hat{S}_{\text{m}}\left[g_{\mu\nu}e^{2\alpha(\phi)}, \Gamma^{\mu}{}_{\nu\rho} + \beta(\phi)\delta^{\mu}_{\nu}\phi_{,\rho}, \chi^I\right]\,,
\end{equation}
with two free functions \(\alpha\) and \(\beta\) of the scalar field and an arbitrary set \(\chi^I\) of matter fields. Writing the variation of the matter action as
\begin{equation}\label{eq:matactvar}
\delta S_{\text{m}} = \int_M\left(\frac{1}{2}\Theta^{\mu\nu}\delta g_{\mu\nu} + H_{\mu}{}^{\nu\rho}\delta\Gamma^{\mu}{}_{\nu\rho} + \Phi\delta\phi + \varpi_I\delta\chi^I\right)\sqrt{-g}\dd^4x\,,
\end{equation}
we identify the energy-momentum tensor \(\Theta_{\mu\nu}\), hypermomentum \(H_{\mu}{}^{\nu\rho}\), non-minimal matter coupling \(\Phi\) and matter field equations \(\varpi^I\). It follows from the assumed structure~\eqref{eq:mataction} of the matter action that these are not independent. This can be seen by writing the variation as
\begin{equation}
\begin{split}
\delta\hat{S}_{\text{m}} &= \int_M\left[\frac{1}{2}\hat{\Theta}^{\mu\nu}\delta\left(g_{\mu\nu}e^{2\alpha}\right) + \hat{H}_{\mu}{}^{\nu\rho}\delta\left(\Gamma^{\mu}{}_{\nu\rho} + \beta\delta^{\mu}_{\nu}\phi_{,\rho}\right) + \hat{\varpi}_I\delta\chi^I\right]e^{4\alpha}\sqrt{-g}\dd^4x\\
&= \int_M\left[\frac{1}{2}\hat{\Theta}^{\mu\nu}e^{2\alpha}(\delta g_{\mu\nu} + 2g_{\mu\nu}\alpha'\delta\phi) + \hat{H}_{\mu}{}^{\nu\rho}(\delta\Gamma^{\mu}{}_{\nu\rho} + \beta'\delta^{\mu}_{\nu}\phi_{,\rho}\delta\phi + \beta\delta^{\mu}_{\nu}\delta\phi_{,\rho}) + \hat{\varpi}_I\delta\chi^I\right]e^{4\alpha}\sqrt{-g}\dd^4x\\
&= \int_M\left\{\frac{1}{2}e^{6\alpha}\hat{\Theta}^{\mu\nu}\delta g_{\mu\nu} + e^{4\alpha}\hat{H}_{\mu}{}^{\nu\rho}\delta\Gamma^{\mu}{}_{\nu\rho} + \left[e^{6\alpha}\alpha'\hat{\Theta}^{\mu\nu}g_{\mu\nu} - \beta\lc{\nabla}_{\nu}\left(e^{4\alpha}\hat{H}_{\mu}{}^{\mu\nu}\right)\right]\delta\phi + e^{4\alpha}\hat{\varpi}_I\delta\chi^I\right\}\sqrt{-g}\dd^4x
\end{split}
\end{equation}
after integration by parts, where here and in the remainder of this article we omit the function arguments for brevity, unless they are required for clarity. By comparison with the variation~\eqref{eq:matactvar}, we can thus identify the terms
\begin{equation}\label{eq:matvar}
\Theta^{\mu\nu} = e^{6\alpha}\hat{\Theta}^{\mu\nu}\,, \quad
H_{\mu}{}^{\nu\rho} = e^{4\alpha}\hat{H}_{\mu}{}^{\nu\rho}\,, \quad
\Phi = e^{6\alpha}\alpha'\hat{\Theta}^{\mu\nu}g_{\mu\nu} - \beta\lc{\nabla}_{\nu}\left(e^{4\alpha}\hat{H}_{\mu}{}^{\mu\nu}\right)\,, \quad
\varpi_I = e^{4\alpha}\hat{\varpi}_I\,,
\end{equation}
and so we find the relation
\begin{equation}\label{eq:matscal}
\Phi = \alpha'\Theta - \beta\lc{\nabla}_{\nu}H_{\mu}{}^{\mu\nu}\,,
\end{equation}
where we have defined \(\Theta = \Theta^{\mu\nu}g_{\mu\nu}\). This relation then also enters the gravitational field equations, which we will not discuss here for brevity.

\subsection{General teleparallel gravity}\label{ssec:genaction}
In the most general case, we will study the class of general teleparallel gravity actions defined by
\begin{equation}\label{eq:genaction}
S_G[g_{\mu\nu}, \Gamma^{\mu}{}_{\nu\rho}, \phi] = \frac{1}{2\kappa^2}\int_M\left[-\mathcal{A}(\phi)G + 2\mathcal{B}(\phi)X + 2\mathcal{C}(\phi)U + 2\mathcal{D}(\phi)V + 2\mathcal{E}(\phi)W - 2\kappa^2\mathcal{V}(\phi)\right]\sqrt{-g}\dd^4x\,,
\end{equation}
where we introduced the abbreviations
\begin{equation}
G = 2M^{\mu}{}_{\rho[\mu}M^{\rho\nu}{}_{\nu]}\,, \quad
X = -\frac{1}{2}g^{\mu\nu}\phi_{,\mu}\phi_{,\nu}\,, \quad
U = T_{\mu}{}^{\mu\nu}\phi_{,\nu}\,, \quad
V = Q^{\nu\mu}{}_{\mu}\phi_{,\nu}\,, \quad
W = Q_{\mu}{}^{\mu\nu}\phi_{,\nu}\,,
\end{equation}
and \(\mathcal{A}, \mathcal{B}, \mathcal{C}, \mathcal{D}, \mathcal{E}, \mathcal{V}\) are functions of the scalar field, whose choice determines a particular action within this class. Theories of this type were studied in~\cite{Hohmann:2023rqn}, and generalize a class of theories discussed in~\cite{Hohmann:2022mlc,Heisenberg:2022mbo}. This action is motivated by analogy to the well-known class of scalar-curvature gravity theories~\cite{Faraoni:2004pi,Fujii:2003pa}. Note that there are several equivalent possibilities to impose the flatness of the connection: either by restricting the variation of the connection, or by introducing a Lagrange multiplier~\cite{Hohmann:2021fpr,Hohmann:2022mlc}. Here we choose the latter approach, and introduce another contribution to the action given by
\begin{equation}\label{eq:curlagmul}
S_{\mathfrak{r}}[\mathfrak{r}_{\mu}{}^{\nu\rho\sigma}, \Gamma^{\mu}{}_{\nu\rho}] = \int_M\mathfrak{r}_{\mu}{}^{\nu\rho\sigma}R^{\mu}{}_{\nu\rho\sigma}\dd^4x\,,
\end{equation}
where the tensor density \(\mathfrak{r}_{\mu}{}^{\nu\rho\sigma}\) is the Lagrange multiplier which enforces the flatness~\eqref{eq:curvature}. The full action of general scalar-teleparallel gravity is then given by
\begin{equation}
S_{\text{gen}} = S_G + S_{\mathfrak{r}} + S_{\text{m}}\,,
\end{equation}
and its properties will be discussed in the following sections.

\subsection{Symmetric teleparallel gravity}\label{ssec:symaction}
In symmetric teleparallel gravity, one considers a connection with vanishing torsion, in addition to the vanishing curvature condition. Following the same line of thought as for the general teleparallel case discussed above, we implement this additional constraint by introducing another Lagrange multiplier term
\begin{equation}
S_{\mathfrak{t}}[\mathfrak{t}_{\mu}{}^{\nu\rho}, \Gamma^{\mu}{}_{\nu\rho}] = \int_M\mathfrak{t}_{\mu}{}^{\nu\rho}T^{\mu}{}_{\nu\rho}\dd^4x
\end{equation}
into the action, with another tensor density \(\mathfrak{t}_{\mu}{}^{\nu\rho}\). Imposing this constraint, we find that the gravity scalar \(G\) reduces to
\begin{equation}
Q = \frac{1}{4}Q^{\mu\nu\rho}Q_{\mu\nu\rho} - \frac{1}{2}Q^{\mu\nu\rho}Q_{\rho\mu\nu} - \frac{1}{4}Q^{\rho\mu}{}_{\mu}Q_{\rho\nu}{}^{\nu} + \frac{1}{2}Q^{\mu}{}_{\mu\rho}Q^{\rho\nu}{}_{\nu}\,,
\end{equation}
while \(U\) vanishes. It follows that the general scalar-teleparallel action~\eqref{eq:genaction} reduces to the scalar-nonmetricity class of actions given by
\begin{equation}\label{eq:symaction}
S_Q[g_{\mu\nu}, \Gamma^{\mu}{}_{\nu\rho}, \phi] = \frac{1}{2\kappa^2}\int_M\left[-\mathcal{A}(\phi)Q + 2\mathcal{B}(\phi)X + 2\mathcal{D}(\phi)V + 2\mathcal{E}(\phi)W - 2\kappa^2\mathcal{V}(\phi)\right]\sqrt{-g}\dd^4x\,,
\end{equation}
which generalizes a class of theories discussed in~\cite{Jarv:2018bgs,Runkla:2018xrv}. The total action is thus given by
\begin{equation}
S_{\text{sym}} = S_Q + S_{\mathfrak{r}} + S_{\mathfrak{t}} + S_{\text{m}}\,,
\end{equation}
and will also be studied in this article.

\subsection{Metric teleparallel gravity}\label{ssec:metaction}
Finally, we come to the metric teleparallel class of theories, where in addition to vanishing curvature we impose vanishing nonmetricity by a Lagrange multiplier term
\begin{equation}
S_{\mathfrak{q}}[\mathfrak{q}^{\mu\nu\rho}, g_{\mu\nu}, \Gamma^{\mu}{}_{\nu\rho}] = \int_M\mathfrak{q}^{\mu\nu\rho}Q_{\mu\nu\rho}\dd^4x\,,
\end{equation}
where \(\mathfrak{q}^{\mu\nu\rho}\) is again a tensor density. Under this constraint, the gravity scalar \(G\) reduces to
\begin{equation}
T = \frac{1}{4}T^{\mu\nu\rho}T_{\mu\nu\rho} + \frac{1}{2}T^{\mu\nu\rho}T_{\rho\nu\mu} - T^{\mu}{}_{\mu\rho}T_{\nu}{}^{\nu\rho}\,,
\end{equation}
while \(V\) and \(W\) vanish. The general scalar-teleparallel action~\eqref{eq:genaction} thus reduces to the scalar-torsion class
\begin{equation}\label{eq:metaction}
S_T[g_{\mu\nu}, \Gamma^{\mu}{}_{\nu\rho}, \phi] = \frac{1}{2\kappa^2}\int_M\left[-\mathcal{A}(\phi)T + 2\mathcal{B}(\phi)X + 2\mathcal{C}(\phi)U - 2\kappa^2\mathcal{V}(\phi)\right]\sqrt{-g}\dd^4x\,,
\end{equation}
which was discussed in detail in~\cite{Hohmann:2018ijr}. The final structure of the action is thus given by
\begin{equation}
S_{\text{met}} = S_T + S_{\mathfrak{r}} + S_{\mathfrak{q}} + S_{\text{m}}\,,
\end{equation}
which resembles that of the previously introduced cases. This concludes the family of scalar-teleparallel gravity theories we study in this work.

\section{Transformation of gravity theories}\label{sec:actiontrans}
We will now study the behavior of the different classes of scalar-teleparallel gravity theories under the transformations of the fundamental fields which we displayed in section~\ref{sec:geom}. For this purpose, we consider a new action functional \(\bar{S}\) for the transformed fields \(\bar{g}_{\mu\nu}, \bar{\Gamma}^{\mu}{}_{\nu\rho}, \bar{\phi}\), which is defined in full analogy to \(S\), but with the parameter functions \(\mathcal{A}, \mathcal{B}, \mathcal{C}, \mathcal{D}, \mathcal{E}, \mathcal{V}, \alpha, \beta\) replaced by new functions \(\bar{\mathcal{A}}, \bar{\mathcal{B}}, \bar{\mathcal{C}}, \bar{\mathcal{D}}, \bar{\mathcal{E}}, \bar{\mathcal{V}}, \bar{\alpha}, \bar{\beta}\). This new action will involve the quantities
\begin{subequations}\label{eq:grascaltra}
\begin{align}
\bar{G} &= e^{-2\gamma}[G + 2\gamma'(2U - V + W) + 12\gamma'^2X]\,,\\
\bar{T} &= e^{-2\gamma}(T + 4\zeta U + 12\zeta^2X)\,,\\
\bar{Q} &= e^{-2\gamma}[Q + 2(\zeta - \gamma')(V - W) + 12(\zeta - \gamma')^2X]\,,\\
\bar{X} &= e^{-2\gamma}f'^2X\,,\\
\bar{U} &= e^{-2\gamma}f'(U + 6\zeta X)\,,\\
\bar{V} &= e^{-2\gamma}f'[V + 16(\zeta - \gamma')X]\,,\\
\bar{W} &= e^{-2\gamma}f'[W + 4(\zeta - \gamma')X]
\end{align}
\end{subequations}
in place of the corresponding unbarred quantities in the gravitational part of the action. As in the previous section, where we outlined the different classes on scalar-teleparallel action we study here, we will proceed for each class separately, starting with the general matter action in section~\ref{ssec:matacttra}. We then continue with the general teleparallel case in section~\ref{ssec:genacttra}, followed by the symmetric teleparallel case in section~\ref{ssec:symacttra} and finally the metric teleparallel case in section~\ref{ssec:metacttra}.

\subsection{Matter part}\label{ssec:matacttra}
We start with the matter part of the new action, which is given by
\begin{equation}\label{eq:matacttra}
\begin{split}
\bar{S}_{\text{m}}\left[\bar{g}_{\mu\nu}, \bar{\Gamma}^{\mu}{}_{\nu\rho}, \bar{\phi}, \chi^I\right] &= \hat{S}_{\text{m}}\left[\bar{g}_{\mu\nu}e^{2\bar{\alpha}(\bar{\phi})}, \bar{\Gamma}^{\mu}{}_{\nu\rho} + \bar{\beta}(\bar{\phi})\delta^{\mu}_{\nu}\bar{\phi}_{,\rho}, \chi^I\right]\\
&= \hat{S}_{\text{m}}\left[g_{\mu\nu}e^{2\bar{\alpha}(f(\phi)) + 2\gamma(\phi)}, \Gamma^{\mu}{}_{\nu\rho} + \zeta(\phi)\delta^{\mu}_{\nu}\phi_{,\rho} + \bar{\beta}(f(\phi))f'(\phi)\delta^{\mu}_{\nu}\phi_{,\rho}, \chi^I\right]\,.
\end{split}
\end{equation}
By comparison with the original action~\eqref{eq:mataction} we now see that both actions for the original fields \(g_{\mu\nu}, \Gamma^{\mu}{}_{\nu\rho}, \phi\) agree if and only if the defining functions are related by
\begin{equation}\label{eq:matfuntra}
\alpha = \bar{\alpha} + \gamma\,, \quad
\beta = f'\bar{\beta} + \zeta\,,
\end{equation}
where we omitted the arguments again, and it is understood that the transformed (barred) functions depend on \(\bar{\phi} = f(\phi)\), while the original (unbarred) functions depend on \(\phi\). Writing the variation as
\begin{equation}
\begin{split}
\delta\bar{S}_{\text{m}} &= \int_M\left(\frac{1}{2}\bar{\Theta}^{\mu\nu}\delta\bar{g}_{\mu\nu} + \bar{H}_{\mu}{}^{\nu\rho}\delta\bar{\Gamma}^{\mu}{}_{\nu\rho} + \bar{\Phi}\delta\bar{\phi} + \bar{\varpi}_I\delta\chi^I\right)\sqrt{-\bar{g}}\dd^4x\\
&= \int_M\left[\frac{1}{2}\bar{\Theta}^{\mu\nu}e^{2\gamma}(\delta g_{\mu\nu} + 2g_{\mu\nu}\gamma'\delta\phi) + \bar{H}_{\mu}{}^{\nu\rho}(\delta\Gamma^{\mu}{}_{\nu\rho} + \zeta'\delta^{\mu}_{\nu}\phi_{,\rho}\delta\phi + \zeta\delta^{\mu}_{\nu}\delta\phi_{,\rho}) + \bar{\Phi}f'\delta\phi + \bar{\varpi}_I\delta\chi^I\right]e^{4\gamma}\sqrt{-g}\dd^4x\\
&= \int_M\left\{\frac{1}{2}e^{6\gamma}\bar{\Theta}^{\mu\nu}\delta g_{\mu\nu} + e^{4\gamma}\bar{H}_{\mu}{}^{\nu\rho}\delta\Gamma^{\mu}{}_{\nu\rho} + \left[e^{4\gamma}\gamma'\bar{\Theta}^{\mu\nu}\bar{g}_{\mu\nu} - \zeta\lc{\nabla}_{\nu}\left(e^{4\gamma}\bar{H}_{\mu}{}^{\mu\nu}\right) + e^{4\gamma}f'\bar{\Phi}\right]\delta\phi + e^{4\gamma}\bar{\varpi}_I\delta\chi^I\right]\sqrt{-g}\dd^4x\,,
\end{split}
\end{equation}
we see that the matter terms transform as
\begin{equation}\label{eq:matvartra}
\Theta^{\mu\nu} = e^{6\gamma}\bar{\Theta}^{\mu\nu}\,, \quad
H_{\mu}{}^{\nu\rho} = e^{4\gamma}\bar{H}_{\mu}{}^{\nu\rho}\,, \quad
\Phi = e^{4\gamma}\left(\gamma'\bar{\Theta} - \zeta\lc{\bar{\nabla}}_{\nu}\bar{H}_{\mu}{}^{\mu\nu} + f'\bar{\Phi}\right)\,, \quad
\varpi_I = e^{4\gamma}\bar{\varpi}_I\,.
\end{equation}
One can perform a few consistency checks on this result. First, note that from the definition of the matter action \(\bar{S}_{\text{m}}\) by the first line of~\eqref{eq:matacttra}, which is analogous to the definition~\eqref{eq:mataction} of \(S_{\text{m}}\), one finds the corresponding relation
\begin{equation}\label{eq:matvar2}
\bar{\Theta}^{\mu\nu} = e^{6\bar{\alpha}}\hat{\Theta}^{\mu\nu}\,, \quad
\bar{H}_{\mu}{}^{\nu\rho} = e^{4\bar{\alpha}}\hat{H}_{\mu}{}^{\nu\rho}\,, \quad
\bar{\Phi} = e^{6\bar{\alpha}}\bar{\alpha}'\hat{\Theta}^{\mu\nu}\bar{g}_{\mu\nu} - \bar{\beta}\lc{\bar{\nabla}}_{\nu}\left(e^{4\bar{\alpha}}\hat{H}_{\mu}{}^{\mu\nu}\right)\,, \quad
\bar{\varpi}_I = e^{4\bar{\alpha}}\hat{\varpi}_I\,,
\end{equation}
from which then also follows
\begin{equation}\label{eq:matscal2}
\bar{\Phi} = \bar{\alpha}'\bar{\Theta} - \bar{\beta}\lc{\bar{\nabla}}_{\nu}\bar{H}_{\mu}{}^{\mu\nu}\,.
\end{equation}
Substituting the matter terms~\eqref{eq:matvar2} in the transformation~\eqref{eq:matvartra} and comparing with the corresponding terms~\eqref{eq:matvar}, one finds again the transformation~\eqref{eq:matfuntra}. Similarly, one finds that also the relations~\eqref{eq:matscal} and~\eqref{eq:matscal2} are connected by the same transformation rules.

\subsection{General teleparallel gravity}\label{ssec:genacttra}
We proceed in full analogy with the gravitational part of the action, starting with the general teleparallel case. With the transformations~\eqref{eq:grascaltra} at hand, we can now replace the action~\eqref{eq:genaction} by the new action
\begin{equation}\label{eq:genacttra}
\begin{split}
\bar{S}_G[\bar{g}_{\mu\nu}, \bar{\Gamma}^{\mu}{}_{\nu\rho}, \bar{\phi}] &= \frac{1}{2\kappa^2}\int_M\left[-\bar{\mathcal{A}}(\bar{\phi})\bar{G} + 2\bar{\mathcal{B}}(\bar{\phi})\bar{X} + 2\bar{\mathcal{C}}(\bar{\phi})\bar{U} + 2\bar{\mathcal{D}}(\bar{\phi})\bar{V} + 2\bar{\mathcal{E}}(\bar{\phi})\bar{W} - 2\kappa^2\bar{\mathcal{V}}(\bar{\phi})\right]\sqrt{-\bar{g}}\dd^4x\\
&= \frac{1}{2\kappa^2}\int_Me^{2\gamma}\bigg\{-\bar{\mathcal{A}}[G + 2\gamma'(2U - V + W) + 12\gamma'^2X] + 2\bar{\mathcal{B}}f'^2X + 2\bar{\mathcal{C}}f'(U + 6\zeta X)\\
&\phantom{=}+ 2\bar{\mathcal{D}}f'[V + 16(\zeta - \gamma')X] + 2\bar{\mathcal{E}}f'[W + 4(\zeta - \gamma')X] - 2\kappa^2e^{2\gamma}\bar{\mathcal{V}}\bigg\}\sqrt{-g}\dd^4x\\
&= \frac{1}{2\kappa^2}\int_Me^{2\gamma}\bigg\{-\bar{\mathcal{A}}G + 2[f'^2\bar{\mathcal{B}} - 6\gamma'^2\bar{\mathcal{A}} + 6\zeta f'\bar{\mathcal{C}} + 4(\zeta - \gamma')f'(4\bar{\mathcal{D}} + \bar{\mathcal{E}})]X\\
&\phantom{=}+ 2(f'\bar{\mathcal{C}} - 2\gamma'\bar{\mathcal{A}})U + 2(f'\bar{\mathcal{D}} + \gamma'\bar{\mathcal{A}})V + 2(f'\bar{\mathcal{E}} - \gamma'\bar{\mathcal{A}})W - 2\kappa^2e^{2\gamma}\bar{\mathcal{V}}\bigg\}\sqrt{-g}\dd^4x\,.
\end{split}
\end{equation}
We see that this reproduces the original action \(S_G\) for the original fields if and only if the parameter functions are related by
\begin{subequations}\label{eq:genfuntra}
\begin{align}
\mathcal{A} &= e^{2\gamma}\bar{\mathcal{A}}\,,\\
\mathcal{B} &= e^{2\gamma}[f'^2\bar{\mathcal{B}} - 6\gamma'^2\bar{\mathcal{A}} + 6\zeta f'\bar{\mathcal{C}} + 4(\zeta - \gamma')f'(4\bar{\mathcal{D}} + \bar{\mathcal{E}})]\,,\\
\mathcal{C} &= e^{2\gamma}(f'\bar{\mathcal{C}} - 2\gamma'\bar{\mathcal{A}})\,,\\
\mathcal{D} &= e^{2\gamma}(f'\bar{\mathcal{D}} + \gamma'\bar{\mathcal{A}})\,,\\
\mathcal{E} &= e^{2\gamma}(f'\bar{\mathcal{E}} - \gamma'\bar{\mathcal{A}})\,,\\
\mathcal{V} &= e^{4\gamma}\bar{\mathcal{V}}\,.
\end{align}
\end{subequations}
In order to complete the transformation of the action, we also need to replace the Lagrange multiplier part by
\begin{equation}\label{eq:curlagtra}
\bar{S}_{\mathfrak{r}}[\bar{\mathfrak{r}}_{\mu}{}^{\nu\rho\sigma}, \bar{\Gamma}^{\mu}{}_{\nu\rho}] = \int_M\bar{\mathfrak{r}}_{\mu}{}^{\nu\rho\sigma}\bar{R}^{\mu}{}_{\nu\rho\sigma}\dd^4x = \int_M\bar{\mathfrak{r}}_{\mu}{}^{\nu\rho\sigma}R^{\mu}{}_{\nu\rho\sigma}\dd^4x\,,
\end{equation}
recalling from the transformation~\eqref{eq:curtrans} that the curvature tensor is invariant under the class of transformations we consider. It follows that this action becomes equal to \(S_{\mathfrak{r}}\) if we set
\begin{equation}
\mathfrak{r}_{\mu}{}^{\nu\rho\sigma} = \bar{\mathfrak{r}}_{\mu}{}^{\nu\rho\sigma}\,.
\end{equation}
In summary, we thus find that under a transformation of the dynamical variables, the total action \(S_{\text{gen}}\) retains its form, where the defining functions obey the transformations~\eqref{eq:matfuntra} and~\eqref{eq:genfuntra}.

\subsection{Symmetric teleparallel gravity}\label{ssec:symacttra}
We then come to the symmetric teleparallel case. Recall from section~\ref{ssec:symaction} that we constructed the action for scalar-nonmetricity gravity by introducing a Lagrange multiplier which imposes vanishing torsion in the scalar-teleparallel action. Here we follow the same procedure to study the behavior of the scalar-nonmetricity theory under field transformations, and start with the Lagrange multiplier part \(S_{\mathfrak{t}}\). Proceeding analogously to the previously discussed actions, we find that it obeys the transformation
\begin{equation}
\bar{S}_{\mathfrak{t}}[\bar{\mathfrak{t}}_{\mu}{}^{\nu\rho}, \bar{\Gamma}^{\mu}{}_{\nu\rho}] = \int_M\bar{\mathfrak{t}}_{\mu}{}^{\nu\rho}\bar{T}^{\mu}{}_{\nu\rho}\dd^4x = \int_M\bar{\mathfrak{t}}_{\mu}{}^{\nu\rho}\left(T^{\mu}{}_{\nu\rho} - 2\zeta\delta^{\mu}_{[\nu}\phi_{,\rho]}\right)\dd^4x\,.
\end{equation}
Hence, we find that this term is, in general, not form-invariant, unless \(\zeta \equiv 0\). This leaves us with two options: most obviously, we could restrict ourselves to transformations which leave this action form-invariant; such transformations, which would be pure conformal transformations of the metric and redefinitions of the scalar field, have been introduced in~\cite{Jarv:2018bgs}. Alternatively, we can study a more general class of actions, which retains its form under the full class of transformations. Here we decide for the latter, and consider a more general Lagrange multiplier term defined by
\begin{equation}\label{eq:torlagmul}
S_{\mathfrak{t}}'[\mathfrak{t}_{\mu}{}^{\nu\rho}, \Gamma^{\mu}{}_{\nu\rho}, \phi] = \int_M\mathfrak{t}_{\mu}{}^{\nu\rho}\left[T^{\mu}{}_{\nu\rho} - 2\mathcal{T}(\phi)\delta^{\mu}_{[\nu}\phi_{,\rho]}\right]\dd^4x\,,
\end{equation}
with another free function \(\mathcal{T}\) of the scalar field. It is now straightforward to calculate the transformation
\begin{equation}
\bar{S}_{\mathfrak{t}}'[\bar{\mathfrak{t}}_{\mu}{}^{\nu\rho}, \bar{\Gamma}^{\mu}{}_{\nu\rho}, \bar{\phi}] = \int_M\bar{\mathfrak{t}}_{\mu}{}^{\nu\rho}\left[\bar{T}^{\mu}{}_{\nu\rho} - 2\bar{\mathcal{T}}(\bar{\phi})\delta^{\mu}_{[\nu}\bar{\phi}_{,\rho]}\right]\dd^4x = \int_M\bar{\mathfrak{t}}_{\mu}{}^{\nu\rho}\left[T^{\mu}{}_{\nu\rho} - 2(f'\bar{\mathcal{T}} + \zeta)\delta^{\mu}_{[\nu}\phi_{,\rho]}\right]\dd^4x\,.
\end{equation}
We see that this indeed reproduces the original action for the original fields, provided that the Lagrange multiplier and the newly introduced function \(\mathcal{T}\) follow the transformation
\begin{equation}\label{eq:symcontra}
\mathfrak{t}_{\mu}{}^{\nu\rho} = \bar{\mathfrak{t}}_{\mu}{}^{\nu\rho}\,, \quad
\mathcal{T} = f'\bar{\mathcal{T}} + \zeta\,.
\end{equation}
However, the enhanced form-invariance comes at a price: the connection is no longer imposed to be torsion-free, but instead the torsion is fixed by the algebraic (i.e., no derivatives on the torsion) constraint
\begin{equation}\label{eq:symconstr}
T^{\mu}{}_{\nu\rho} = 2\mathcal{T}(\phi)\delta^{\mu}_{[\nu}\phi_{,\rho]}\,,
\end{equation}
which is non-vanishing in general. It follows that under this constraint the general scalar-teleparallel action \(S_G\)~\eqref{eq:genaction} does not reduce to the scalar-nonmetricity action \(S_Q\)~\eqref{eq:symaction}. Imposing the constraint~\eqref{eq:symconstr}, we find that the scalar terms in the action become
\begin{equation}
G = Q + 2\mathcal{T}(V - W) + 12\mathcal{T}^2X\,, \quad
U = -6\mathcal{T}X\,.
\end{equation}
It thus follows that under this constraint the general scalar-teleparallel action~\eqref{eq:genaction} becomes equivalent to
\begin{equation}
\begin{split}
S_Q'[g_{\mu\nu}, \Gamma^{\mu}{}_{\nu\rho}, \phi] &= \frac{1}{2\kappa^2}\int_M\left\{-\mathcal{A}[Q + 2\mathcal{T}(V - W) + 12\mathcal{T}^2X] + 2\mathcal{B}X - 12\mathcal{C}\mathcal{T}X + 2\mathcal{D}V + 2\mathcal{E}W - 2\kappa^2\mathcal{V}\right\}\sqrt{-g}\dd^4x\\
&= \frac{1}{2\kappa^2}\int_M\left[-\mathcal{A}Q + 2(\mathcal{B} - 6\mathcal{C}\mathcal{T} - 6\mathcal{A}\mathcal{T}^2)X + 2(\mathcal{D} - \mathcal{A}\mathcal{T})V + 2(\mathcal{E} + \mathcal{A}\mathcal{T})W - 2\kappa^2\mathcal{V}\right]\sqrt{-g}\dd^4x\\
&= \frac{1}{2\kappa^2}\int_M\left(-\st{\mathcal{A}}Q + 2\st{\mathcal{B}}X + 2\st{\mathcal{D}}V + 2\st{\mathcal{E}}W - 2\kappa^2\st{\mathcal{V}}\right)\sqrt{-g}\dd^4x\,.
\end{split}
\end{equation}
Note that formally this action \(S_Q'\) has the same form as \(S_Q\), but with different parameter functions
\begin{equation}\label{eq:symactfunc}
\st{\mathcal{A}} = \mathcal{A}\,, \quad
\st{\mathcal{B}} = \mathcal{B} - 6\mathcal{C}\mathcal{T} - 6\mathcal{A}\mathcal{T}^2\,, \quad
\st{\mathcal{D}} = \mathcal{D} - \mathcal{A}\mathcal{T}\,, \quad
\st{\mathcal{E}} = \mathcal{E} + \mathcal{A}\mathcal{T}\,, \quad
\st{\mathcal{V}} = \mathcal{V}\,,
\end{equation}
which agree with the original functions only if \(\mathcal{T} \equiv 0\). Hence, in the following we will study the class of theories defined by the generalized scalar-nonmetricity action
\begin{equation}
S_{\text{sym}}' = S_Q' + S_{\mathfrak{r}} + S_{\mathfrak{t}}' + S_{\text{m}}\,.
\end{equation}
We can then proceed by studying the transformation of the action \(S_Q'\) under field transformations, and we find that this is given by
\begin{equation}\label{eq:symacttra}
\begin{split}
\bar{S}_Q'[\bar{g}_{\mu\nu}, \bar{\Gamma}^{\mu}{}_{\nu\rho}, \bar{\phi}] &= \frac{1}{2\kappa^2}\int_M\left[-\st{\bar{\mathcal{A}}}(\bar{\phi})\bar{Q} + 2\st{\bar{\mathcal{B}}}(\bar{\phi})\bar{X} + 2\st{\bar{\mathcal{D}}}(\bar{\phi})\bar{V} + 2\st{\bar{\mathcal{E}}}(\bar{\phi})\bar{W} - 2\kappa^2\st{\bar{\mathcal{V}}}(\bar{\phi})\right]\sqrt{-\bar{g}}\dd^4x\\
&= \frac{1}{2\kappa^2}\int_Me^{2\gamma}\bigg\{-\st{\bar{\mathcal{A}}}[Q + 2(\zeta - \gamma')(V + W) + 12(\zeta - \gamma')^2X] + 2\st{\bar{\mathcal{B}}}f'^2X\\
&\phantom{=}+ 2\st{\bar{\mathcal{D}}}f'[V + 16(\zeta - \gamma')X] + 2\st{\bar{\mathcal{E}}}f'[W + 4(\zeta - \gamma')X] - 2\kappa^2e^{2\gamma}\st{\bar{\mathcal{V}}}\bigg\}\sqrt{-g}\dd^4x\\
&= \frac{1}{2\kappa^2}\int_Me^{2\gamma}\bigg\{-\st{\bar{\mathcal{A}}}Q + 2\left[f'^2\st{\bar{\mathcal{B}}} - 6(\zeta - \gamma')^2\st{\bar{\mathcal{A}}} + 4(\zeta - \gamma')f'\left(4\st{\bar{\mathcal{D}}} + \st{\bar{\mathcal{E}}}\right)\right]X\\
&\phantom{=}+ 2\left[f'\st{\bar{\mathcal{D}}} - (\zeta - \gamma')\st{\bar{\mathcal{A}}}\right]V + 2\left[f'\st{\bar{\mathcal{E}}} + (\zeta - \gamma')\st{\bar{\mathcal{A}}}\right]W - 2\kappa^2e^{2\gamma}\st{\bar{\mathcal{V}}}\bigg\}\sqrt{-g}\dd^4x\,.
\end{split}
\end{equation}
We thus find that this action retains its form, provided that the parameter functions undergo the transformations
\begin{subequations}\label{eq:symfuntra}
\begin{align}
\st{\mathcal{A}} &= e^{2\gamma}\st{\bar{\mathcal{A}}}\,,\\
\st{\mathcal{B}} &= e^{2\gamma}\left[f'^2\st{\bar{\mathcal{B}}} - 6(\zeta - \gamma')^2\st{\bar{\mathcal{A}}} + 4(\zeta - \gamma')f'\left(4\st{\bar{\mathcal{D}}} + \st{\bar{\mathcal{E}}}\right)\right]\,,\\
\st{\mathcal{D}} &= e^{2\gamma}\left[f'\st{\bar{\mathcal{D}}} - (\zeta - \gamma')\st{\bar{\mathcal{A}}}\right]\,,\\
\st{\mathcal{E}} &= e^{2\gamma}\left[f'\st{\bar{\mathcal{E}}} + (\zeta - \gamma')\st{\bar{\mathcal{A}}}\right]\,,\\
\st{\mathcal{V}} &= e^{4\gamma}\st{\bar{\mathcal{V}}}\,.
\end{align}
\end{subequations}
We see that these transformations apparently differ from the transformations~\eqref{eq:genfuntra} we have found in the general teleparallel case. Nevertheless, they are closely related, and in fact equivalent: if we make use of their definition~\eqref{eq:symactfunc} for both the original and the transformed functions, and then apply the transformations~\eqref{eq:genfuntra} and~\eqref{eq:symcontra}, we obtain again the transformation rules~\eqref{eq:symfuntra}. We also remark that \(\zeta\) appears only in the combination \(\zeta - \gamma'\), which originates from the transformation~\eqref{eq:nomtrans} of the nonmetricity. This will be used later in this article.

\subsection{Metric teleparallel gravity}\label{ssec:metacttra}
We finally come to the metric teleparallel case, which we discuss following the same steps as in the symmetric teleparallel case studied above. In this case, we have constructed a scalar-torsion action in section~\ref{ssec:metaction} by introducing a Lagrange multiplier term \(S_{\mathfrak{q}}\), in order to impose vanishing nonmetricity. Under a field transformation, we find that this term transforms as
\begin{equation}
\bar{S}_{\mathfrak{q}}[\bar{\mathfrak{q}}^{\mu\nu\rho}, \bar{g}_{\mu\nu}, \bar{\Gamma}^{\mu}{}_{\nu\rho}] = \int_M\bar{\mathfrak{q}}^{\mu\nu\rho}\bar{Q}_{\mu\nu\rho}\dd^4x = \int_M\bar{\mathfrak{q}}^{\mu\nu\rho}\left[Q_{\mu\nu\rho} - 2(\zeta - \gamma')g_{\nu\rho}\phi_{,\mu}\right]e^{2\gamma}\dd^4x\,.
\end{equation}
Again we see that this part of the action does not retain its form unless \(\zeta \equiv \gamma'\); transformations which obey this restriction have been discussed in detail in~\cite{Hohmann:2018ijr}. Here, however, we aim to lift this restriction, and enlarge the class of theories such that it retains its form under the general class of field transformations we study in this article. For this purpose, we modify the Lagrange multiplier term such that it reads
\begin{equation}\label{eq:nomlagmul}
S_{\mathfrak{q}}'[\mathfrak{q}^{\mu\nu\rho}, g_{\mu\nu}, \Gamma^{\mu}{}_{\nu\rho}, \phi] = \int_M\mathfrak{q}^{\mu\nu\rho}\left[Q_{\mu\nu\rho} - 2\mathcal{Q}(\phi)g_{\nu\rho}\phi_{,\mu}\right]\dd^4x\,,
\end{equation}
with a newly introduced function \(\mathcal{Q}\) of the scalar field. This modified term undergoes the transformation
\begin{equation}
\bar{S}_{\mathfrak{q}}'[\bar{\mathfrak{q}}^{\mu\nu\rho}, \bar{g}_{\mu\nu}, \bar{\Gamma}^{\mu}{}_{\nu\rho}, \bar{\phi}] = \int_M\bar{\mathfrak{q}}^{\mu\nu\rho}\left[\bar{Q}_{\mu\nu\rho} - 2\bar{\mathcal{Q}}(\bar{\phi})\bar{g}_{\nu\rho}\bar{\phi}_{,\mu}\right]\dd^4x = \int_M\bar{\mathfrak{q}}^{\mu\nu\rho}\left[Q_{\mu\nu\rho} - 2(f'\bar{\mathcal{Q}} + \zeta - \gamma')g_{\nu\rho}\phi_{,\mu}\right]e^{2\gamma}\dd^4x\,,
\end{equation}
and so we see that it resembles the original term if and only if we set
\begin{equation}\label{eq:metcontra}
\mathfrak{q}^{\mu\nu\rho} = e^{2\gamma}\bar{\mathfrak{q}}^{\mu\nu\rho}\,, \quad
\mathcal{Q} = f'\bar{\mathcal{Q}} + \zeta - \gamma'\,.
\end{equation}
The price we have to pay for this enhanced form-invariance, besides introducing another free function \(\mathcal{Q}\) into the action, is the fact that the connection is no longer metric-compatible, but now possesses non-vanishing nonmetricity
\begin{equation}\label{eq:metconstr}
Q_{\mu\nu\rho} = 2\mathcal{Q}(\phi)g_{\nu\rho}\,.
\end{equation}
We can view this condition as an algebraic constraint on the teleparallel connection coefficients \(\Gamma^{\mu}{}_{\nu\rho}\). Imposing this constraint, we find that the general scalar-teleparallel action \(S_G\)~\eqref{eq:genaction} does not reduce to the scalar-torsion action \(S_T\)~\eqref{eq:metaction} anymore. Note that under the constraint~\eqref{eq:metconstr} the scalar terms in the action reduce to
\begin{equation}
G = T + 4\mathcal{Q}U + 12\mathcal{Q}^2X\,, \quad
V = -16\mathcal{Q}X\,, \quad
W = -4\mathcal{Q}X\,.
\end{equation}
Inserting these relations in the general scalar-teleparallel action~\eqref{eq:genaction}, we find that it becomes equivalent to
\begin{equation}
\begin{split}
S_T'[g_{\mu\nu}, \Gamma^{\mu}{}_{\nu\rho}, \phi] &= \frac{1}{2\kappa^2}\int_M\left[-\mathcal{A}(T + 4\mathcal{Q}U + 12\mathcal{Q}^2X) + 2\mathcal{B}X + 2\mathcal{C}U - 32\mathcal{D}X - 8\mathcal{E}X - 2\kappa^2\mathcal{V}\right]\sqrt{-g}\dd^4x\\
&= \frac{1}{2\kappa^2}\int_M\left[-\mathcal{A}T + 2(\mathcal{B} - 16\mathcal{D}\mathcal{Q} - 4\mathcal{E}\mathcal{Q} - 6\mathcal{A}\mathcal{Q}^2)X + 2(\mathcal{C} - 2\mathcal{A}\mathcal{Q})U - 2\kappa^2\mathcal{V}\right]\sqrt{-g}\dd^4x\\
&= \frac{1}{2\kappa^2}\int_M\left[-\mt{\mathcal{A}}T + 2\mt{\mathcal{B}}X + 2\mt{\mathcal{C}}U - 2\kappa^2\mt{\mathcal{V}}\right]\sqrt{-g}\dd^4x\,,
\end{split}
\end{equation}
which has formally the same form as \(S_T\), but now depends on the parameter functions
\begin{equation}\label{eq:metactfunc}
\mt{\mathcal{A}} = \mathcal{A}\,, \quad
\mt{\mathcal{B}} = \mathcal{B} - 16\mathcal{D}\mathcal{Q} - 4\mathcal{E}\mathcal{Q} - 6\mathcal{A}\mathcal{Q}^2\,, \quad
\mt{\mathcal{C}} = \mathcal{C} - 2\mathcal{A}\mathcal{Q}\,, \quad
\mt{\mathcal{V}} = \mathcal{V}\,.
\end{equation}
The total action we will study in the following is thus given as
\begin{equation}
S_{\text{met}}' = S_T' + S_{\mathfrak{r}} + S_{\mathfrak{q}}' + S_{\text{m}}\,.
\end{equation}
We finally also derive the transformation of \(S_T'\) under general field transformations. In this case, we find the transformation
\begin{equation}\label{eq:metacttra}
\begin{split}
\bar{S}_T'[\bar{g}_{\mu\nu}, \bar{\Gamma}^{\mu}{}_{\nu\rho}, \bar{\phi}] &= \frac{1}{2\kappa^2}\int_M\left[-\mt{\bar{\mathcal{A}}}(\bar{\phi})\bar{T} + 2\mt{\bar{\mathcal{B}}}(\bar{\phi})\bar{X} + 2\mt{\bar{\mathcal{C}}}(\bar{\phi})\bar{U} - 2\kappa^2\mt{\bar{\mathcal{V}}}(\bar{\phi})\right]\sqrt{-\bar{g}}\dd^4x\\
&= \frac{1}{2\kappa^2}\int_Me^{2\gamma}\left[-\mt{\bar{\mathcal{A}}}(T + 4\zeta U + 12\zeta^2X) + 2\mt{\bar{\mathcal{B}}}f'^2X + 2\mt{\bar{\mathcal{C}}}f'(U + 6\zeta X) - 2\kappa^2e^{2\gamma}\mt{\bar{\mathcal{V}}}\right]\sqrt{-g}\dd^4x\\
&= \frac{1}{2\kappa^2}\int_Me^{2\gamma}\left[-\mt{\bar{\mathcal{A}}}T + 2\left(f'^2\mt{\bar{\mathcal{B}}} - 6\zeta^2\mt{\bar{\mathcal{A}}} + 6\zeta f'\mt{\bar{\mathcal{C}}}\right)X + 2\left(f'\mt{\bar{\mathcal{C}}} - 2\zeta\mt{\bar{\mathcal{A}}}\right)U - 2\kappa^2e^{2\gamma}\mt{\bar{\mathcal{V}}}\right]\sqrt{-g}\dd^4x\,.
\end{split}
\end{equation}
It follows that the action remains form-invariant, where the functions in the original and transformed actions are related by
\begin{subequations}\label{eq:metfuntra}
\begin{align}
\mt{\mathcal{A}} &= e^{2\gamma}\mt{\bar{\mathcal{A}}}\,,\\
\mt{\mathcal{B}} &= e^{2\gamma}\left(f'^2\mt{\bar{\mathcal{B}}} - 6\zeta^2\mt{\bar{\mathcal{A}}} + 6\zeta f'\mt{\bar{\mathcal{C}}}\right)\,,\\
\mt{\mathcal{C}} &= e^{2\gamma}\left(f'\mt{\bar{\mathcal{C}}} - 2\zeta\mt{\bar{\mathcal{A}}}\right)\,,\\
\mt{\mathcal{V}} &= e^{4\gamma}\mt{\bar{\mathcal{V}}}\,.
\end{align}
\end{subequations}
We finally remark that we could have obtained the same relations by making use of the definition~\eqref{eq:metactfunc}, together with the transformations~\eqref{eq:genfuntra} and~\eqref{eq:symcontra} of the constituting functions. With these transformations at hand, we conclude the discussion of the transformation of scalar-teleparallel gravity actions, and will study their properties in the following sections.

\section{Invariant quantities}\label{sec:invariant}
We have seen in section~\ref{sec:actiontrans} that the actions of the different classes of scalar-teleparallel gravity theories which we study in this article retain their form under the given class of field transformations of the metric, connection and scalar field, provided that we perform a suitable transformation of the free functions \(\mathcal{A}, \mathcal{B}, \mathcal{C}, \mathcal{D}, \mathcal{E}, \mathcal{V}, \mathcal{T}, \mathcal{Q}, \alpha, \beta\) which select a particular action from these classes. A similar behavior is well known from scalar-curvature theories of gravity~\cite{Flanagan:2004bz}, as well as similarly constructed scalar-torsion theories~\cite{Hohmann:2018ijr}. This fact has motivated the construction of a number of invariant quantities in scalar-curvature~\cite{Jarv:2014hma,Jarv:2015kga,Kuusk:2016rso,Jarv:2016sow,Karam:2017zno} and scalar-torsion~\cite{Hohmann:2018ijr} theories, and it has been conjectured that any observables derived from such theories can be expressed completely in terms of these invariants. In this section, we show that similar invariant quantities can also be constructed for general scalar-teleparallel theories as well as their symmetric and metric counterparts.

We start with the functions \(\mathcal{A}\), \(\mathcal{V}\) and \(\alpha\), for which we have seen that their transformation depends on \(\gamma\) only, without any derivatives. For these we can proceed in full analogy to the scalar-curvature and scalar-torsion cases and define
\begin{equation}
\mathcal{I} = \frac{e^{2\alpha}}{\mathcal{A}}\,, \quad
\mathcal{U} = \frac{\mathcal{V}}{\mathcal{A}^2}\,.
\end{equation}
These functions of the scalar field are true invariants in the sense that under an arbitrary field transformation defined by functions \(\gamma, \zeta, f\) they transform as
\begin{equation}
\bar{\mathcal{I}}(\bar{\phi}(x)) = \bar{\mathcal{I}}(f(\phi(x))) = \mathcal{I}(\phi(x))\,,
\end{equation}
and analogously for \(\mathcal{U}\), which means that even though the functional form of \(\mathcal{I}\) as a function of its scalar field changes due to the fact that the scalar field variable changes, this change is such that it simply compensates the transformation of the scalar field, and the value of these invariants remains the same at every spacetime point, independently of the choice of the scalar field variable through which it is evaluated. We then continue with the functions \(\mathcal{C}\), \(\mathcal{D}\) and \(\mathcal{E}\), whose transformation involves also the derivative \(\gamma'\). In order to compensate this term in the transformation, we can combine these functions with a suitable derivative term
\begin{equation}
\mathcal{A}' = e^{2\gamma}(2\gamma'\bar{\mathcal{A}} + f'\bar{\mathcal{A}}')\,, \quad
\alpha' = f'\bar{\alpha}' + \gamma'\,,
\end{equation}
where it is understood that a prime on a barred function denotes its derivative with respect to the function argument \(\bar{\phi}\), which explains the appearance of \(f'\) from the inner derivative. This allows us to construct the invariants
\begin{equation}
\mathcal{K} = \frac{\mathcal{C} + 2\alpha'\mathcal{A}}{2e^{2\alpha}}\,, \quad
\mathcal{M} = \frac{\mathcal{D} - \alpha'\mathcal{A}}{2e^{2\alpha}}\,, \quad
\mathcal{N} = \frac{\mathcal{E} + \alpha'\mathcal{A}}{2e^{2\alpha}}\,,
\end{equation}
or alternatively
\begin{equation}
\mathcal{H} = \frac{\mathcal{C} + \mathcal{A}'}{2\mathcal{A}}\,, \quad
\mathcal{J} = \frac{2\mathcal{D} - \mathcal{A}'}{4\mathcal{A}}\,, \quad
\mathcal{L} = \frac{2\mathcal{E} + \mathcal{A}'}{4\mathcal{A}}\,.
\end{equation}
As for the previously defined quantities \(\mathcal{I}\) and \(\mathcal{U}\), these are invariant under transformations of the metric with an arbitrary function \(\gamma\), and trivially also invariant under connection transformations with a function \(\zeta\). However, under a redefinition of the scalar field they transform covariantly as
\begin{equation}\label{eq:covtra1}
\mathcal{K} = f'\bar{\mathcal{K}}\,,
\end{equation}
and equivalently for the remaining invariants. We proceed with the functions \(\mathcal{T}\) and \(\beta\), whose transformation involves \(\zeta\), but does not involve \(\gamma\). This transformation therefore cannot be compensated by any of the previously discussed functions, but we can define the invariant
\begin{equation}
\mathcal{S} = \mathcal{T} - \beta\,,
\end{equation}
which is invariant under transformations of the metric and the connection, but follows the covariant transformation~\eqref{eq:covtra1}. We then come to \(\mathcal{Q}\), whose transformation involves the combination \(\zeta - \gamma'\). Building upon the construction of other invariants performed above, we find the invariant
\begin{equation}
\mathcal{P} = \mathcal{Q} + \alpha' - \beta\,,
\end{equation}
which again satisfies the covariant transformation rule~\eqref{eq:covtra1}. Finally, we are left with \(\mathcal{B}\), whose transformation behavior is the most involved. We can follow a similar approach as above and introduce compensating terms involving \(\alpha\) and \(\beta\). One then finds that the function
\begin{equation}
\mathcal{G} = \frac{\mathcal{B} - 6\alpha'^2\mathcal{A} - 6\beta\mathcal{C} + 4(\alpha' - \beta)(4\mathcal{D} + \mathcal{E})}{2e^{2\alpha}}
\end{equation}
is invariant under transformations of the metric and the connection, while under scalar field reparametrizations it transforms as
\begin{equation}\label{eq:covtra2}
\mathcal{G} = f'^2\bar{\mathcal{G}}\,.
\end{equation}
Note that the invariants above are not independent; in particular, we have the relations
\begin{equation}
2(\mathcal{I}\mathcal{K} - \mathcal{H}) = -4(\mathcal{I}\mathcal{M} - \mathcal{J}) = 4(\mathcal{I}\mathcal{N} - \mathcal{L}) = \frac{\mathcal{I}'}{\mathcal{I}}\,,
\end{equation}
which hints towards the possibility to construct new invariants from those given above. In particular, one can use the functions appearing in the symmetric and metric teleparallel action functionals to construct the invariants
\begin{equation}
\mt{\mathcal{K}} = \frac{\mt{\mathcal{C}} + 2\beta\mt{\mathcal{A}}}{2e^{2\alpha}}\,, \quad
\st{\mathcal{M}} = \frac{\st{\mathcal{D}} - (\alpha' - \beta)\st{\mathcal{A}}}{2e^{2\alpha}}\,, \quad
\st{\mathcal{N}} = \frac{\st{\mathcal{E}} + (\alpha' - \beta)\st{\mathcal{A}}}{2e^{2\alpha}}\,,
\end{equation}
or alternatively
\begin{equation}
\mt{\mathcal{H}} = \frac{\mt{\mathcal{C}} + \mt{\mathcal{A}}' + 2\mt{\mathcal{A}}\mathcal{Q}}{2\mt{\mathcal{A}}}\,, \quad
\st{\mathcal{J}} = \frac{2\st{\mathcal{D}} - \st{\mathcal{A}}' + 2\st{\mathcal{A}}\mathcal{T}}{4\st{\mathcal{A}}}\,, \quad
\st{\mathcal{L}} = \frac{2\st{\mathcal{E}} + \st{\mathcal{A}}' - 2\st{\mathcal{A}}\mathcal{T}}{4\st{\mathcal{A}}}\,.
\end{equation}
Through the definitions~\eqref{eq:symactfunc} and~\eqref{eq:metactfunc}, they are related to the previously defined invariants by
\begin{equation}
\mt{\mathcal{K}} = \mathcal{K} - \frac{\mathcal{P}}{\mathcal{I}}\,,\quad
\st{\mathcal{M}} = \mathcal{M} - \frac{\mathcal{S}}{2\mathcal{I}}\,,\quad
\st{\mathcal{N}} = \mathcal{N} + \frac{\mathcal{S}}{2\mathcal{I}}\,,\quad
\mt{\mathcal{H}} = \mathcal{H}\,,\quad
\st{\mathcal{J}} = \mathcal{J}\,,\quad
\st{\mathcal{L}} = \mathcal{L}\,,\quad
\end{equation}
and satisfy the relations
\begin{equation}
2\left(\mathcal{I}\mt{\mathcal{K}} - \mt{\mathcal{H}}\right) = \frac{\mathcal{I}'}{\mathcal{I}} - 2\mathcal{P}\,, \quad
-4\left(\mathcal{I}\st{\mathcal{M}} - \st{\mathcal{J}}\right) = 4\left(\mathcal{I}\st{\mathcal{N}} - \st{\mathcal{L}}\right) = \frac{\mathcal{I}'}{\mathcal{I}} + 2\mathcal{S}\,.
\end{equation}
From the quantities \(\st{\mathcal{B}}\) and \(\mt{\mathcal{B}}\) one can similarly define
\begin{equation}
\st{\mathcal{G}} = \frac{\st{\mathcal{B}} - 6(\alpha' - \beta)^2\st{\mathcal{A}} + 4(\alpha' - \beta)(4\st{\mathcal{D}} + \st{\mathcal{E}})}{2e^{2\alpha}}\,, \quad
\mt{\mathcal{G}} = \frac{\mt{\mathcal{B}} - 6\beta^2\mt{\mathcal{A}} - 6\beta\mt{\mathcal{C}}}{2e^{2\alpha}}\,,
\end{equation}
but one also has another possibility, using either \(\st{\mathcal{A}}\) and \(\mathcal{T}\) or \(\mt{\mathcal{A}}\) and \(\mathcal{Q}\) in place of \(\alpha\) and \(\beta\) as compensating terms, which then leads to
\begin{equation}
\st{\mathcal{F}} = \frac{2\st{\mathcal{A}}\st{\mathcal{B}} - 3(\st{\mathcal{A}}' - 2\st{\mathcal{A}}\mathcal{T})^2 + 4(\st{\mathcal{A}}' - 2\st{\mathcal{A}}\mathcal{T})(4\st{\mathcal{D}} + \st{\mathcal{E}})}{4\st{\mathcal{A}}^2}\,, \quad
\mt{\mathcal{F}} = \frac{2\mt{\mathcal{A}}\mt{\mathcal{B}} - 3(\mt{\mathcal{A}}' + 2\mt{\mathcal{A}}\mathcal{Q})^2 - 6(\mt{\mathcal{A}}' + 2\mt{\mathcal{A}}\mathcal{Q})\mt{\mathcal{C}}}{4\mt{\mathcal{A}}^2}\,.
\end{equation}
Again we remark that the former two terms are obtained from previously defined invariants via
\begin{equation}
\st{\mathcal{G}} = \mathcal{G} - 6\mathcal{K}\mathcal{S} - \frac{3\mathcal{S}^2}{\mathcal{I}}\,, \quad
\mt{\mathcal{G}} = \mathcal{G} - 4(4\mathcal{M} + \mathcal{N})\mathcal{P} - \frac{3\mathcal{P}^2}{\mathcal{I}}
\end{equation}
while they are interrelated by
\begin{equation}
\st{\mathcal{F}} = \mathcal{I}\st{\mathcal{G}} - 2(4\st{\mathcal{M}} + \st{\mathcal{N}})(2\mathcal{I}\mathcal{S} + \mathcal{I}') - \frac{3(2\mathcal{I}\mathcal{S} + \mathcal{I}')^2}{4\mathcal{I}^2}\,, \quad
\mt{\mathcal{F}} = \mathcal{I}\mt{\mathcal{G}} - 3\mt{\mathcal{K}}(2\mathcal{I}\mathcal{P} - \mathcal{I}') - \frac{3(2\mathcal{I}\mathcal{P} - \mathcal{I}')^2}{4\mathcal{I}^2}\,.
\end{equation}
So far the quantities we have defined appear rather arbitrary, except for the sole fact that they are invariant under transformations of the metric and the teleparallel affine connection. However, it turns out that the invariant combinations we have shown here are particularly useful for an invariant formulation of scalar-teleparallel gravity theories, as we will see in the following section. Further, also their remaining transformation behavior~\eqref{eq:covtra1} and~\eqref{eq:covtra2} is not by accident, but has a clear geometric meaning, which will become clear in section~\ref{ssec:minvariant}, where we extend the construction of invariants to multiple scalar fields.

\section{Frames}\label{sec:frames}
The full virtue of the invariant quantities we have defined in the previous section lies in the fact that they can be used for a formulation of scalar-teleparallel gravity theories which is invariant under transformations of the metric and the teleparallel affine connection. The key ingredient to this formulation is the definition of a set of invariant field variables, in terms on which the action and field equations of scalar-teleparallel gravity theories can be expressed with the help of invariant functions only. Here we present two such choices of variables, or frames in a terminology borrowed from scalar-curvature theories. These have properties similar to the well-known Jordan frame, as discussed in section~\ref{ssec:jordan}, as well as the Einstein frame, discussed in section~\ref{ssec:einstein}, which appear in scalar-curvature gravity theories. Since these frames are obtained only through transformations of the metric and the teleparallel affine connection, we will leave the scalar field unchanged, unless stated otherwise.

\subsection{Jordan-like frame}\label{ssec:jordan}
In the scalar-curvature class of gravity theories, the Jordan frame, whose associated quantities we will define with the letter \(\mathfrak{J}\), is defined such that there is no direct coupling between the scalar field and any matter fields~\cite{Flanagan:2004bz}. By comparison with the matter action~\eqref{eq:mataction} we see that this is the case if and only if both coupling functions vanish, hence
\begin{equation}\label{eq:jfcond}
\jf{\alpha} \equiv \jf{\beta} \equiv 0\,.
\end{equation}
By setting the transformed frame, which we denoted with a bar in previous sections, equal to the Jordan frame, we see from the transformation~\eqref{eq:matfuntra} that we can transform from any other frame into the Jordan frame by applying the transformation
\begin{equation}\label{eq:jftrans}
\jf{\gamma} = \alpha\,, \quad
\jf{\zeta} = \beta\,,
\end{equation}
where \(\alpha\) and \(\beta\) are the coupling functions defined in the original frame, from which the transformation is to be performed. One finds that under this transformation the metric and the connection become
\begin{equation}
\jf{g}_{\mu\nu} = e^{2\alpha}g_{\mu\nu}\,, \quad
\jf{\Gamma}^{\mu}{}_{\nu\rho} = \Gamma^{\mu}{}_{\nu\rho} + \beta\delta^{\mu}_{\nu}\phi_{,\rho}\,.
\end{equation}
Note that these field variables are invariant in the sense that if we had started from any other frame, which is related by transformations \(\gamma\) and \(\zeta\), we would find
\begin{equation}
\jf{g}_{\mu\nu} = e^{2\alpha}g_{\mu\nu} = e^{2(\bar{\alpha} + \gamma)}g_{\mu\nu} = e^{2\bar{\alpha}}\bar{g}_{\mu\nu} = \jf{\bar{g}}_{\mu\nu}\,,
\end{equation}
as well as
\begin{equation}
\jf{\Gamma}^{\mu}{}_{\nu\rho} = \Gamma^{\mu}{}_{\nu\rho} + \beta\delta^{\mu}_{\nu}\phi_{,\rho} = \Gamma^{\mu}{}_{\nu\rho} + (\bar{\beta} + \zeta)\delta^{\mu}_{\nu}\phi_{,\rho} = \bar{\Gamma}^{\mu}{}_{\nu\rho} + \bar{\beta}\delta^{\mu}_{\nu}\phi_{,\rho} = \jf{\bar{\Gamma}}^{\mu}{}_{\nu\rho}\,.
\end{equation}
Further, one finds that not only the field variables become invariant, but also the functions in the gravitational action. Applying the transformation~\eqref{eq:jftrans} with the rules~\eqref{eq:genfuntra}, one finds that in the Jordan frame the defining functions become
\begin{equation}\label{eq:jfgenfunc}
\jf{\mathcal{A}} = \frac{1}{\mathcal{I}}\,, \quad
\jf{\mathcal{B}} = 2\mathcal{G}\,, \quad
\jf{\mathcal{C}} = 2\mathcal{K}\,, \quad
\jf{\mathcal{D}} = 2\mathcal{M}\,, \quad
\jf{\mathcal{E}} = 2\mathcal{N}\,, \quad
\jf{\mathcal{V}} = \frac{\mathcal{U}}{\mathcal{I}^2}\,.
\end{equation}
A similar result is obtained for the scalar-nonmetricity class of theories. With the transformation rules~\eqref{eq:symcontra} and~\eqref{eq:symfuntra} one finds
\begin{equation}\label{eq:jfsymfunc}
\jf{\st{\mathcal{A}}} = \frac{1}{\mathcal{I}}\,, \quad
\jf{\st{\mathcal{B}}} = 2\st{\mathcal{G}}\,, \quad
\jf{\st{\mathcal{D}}} = 2\st{\mathcal{M}}\,, \quad
\jf{\st{\mathcal{E}}} = 2\st{\mathcal{N}}\,, \quad
\jf{\st{\mathcal{V}}} = \frac{\mathcal{U}}{\mathcal{I}^2}\,, \quad
\jf{\mathcal{T}} = \mathcal{S}\,.
\end{equation}
Finally, one can apply the same transformation also to scalar-torsion theories, for which the rules~\eqref{eq:metcontra} and~\eqref{eq:metfuntra} yield
\begin{equation}\label{eq:jfmetfunc}
\jf{\mt{\mathcal{A}}} = \frac{1}{\mathcal{I}}\,, \quad
\jf{\mt{\mathcal{B}}} = 2\mt{\mathcal{G}}\,, \quad
\jf{\mt{\mathcal{C}}} = 2\mt{\mathcal{K}}\,, \quad
\jf{\mt{\mathcal{V}}} = \frac{\mathcal{U}}{\mathcal{I}^2}\,, \quad
\jf{\mathcal{Q}} = \mathcal{P}\,.
\end{equation}
In summary, we see that in the Jordan frame any scalar-teleparallel gravity theory which belongs to one of the three classes we discuss here is characterized by a number of invariants which appear in the gravitational part of the action only. Note that one still has the freedom to reparametrize the scalar field \(\phi\) with an arbitrary invertible function \(f\), and that only \(\mathcal{I}\) and \(\mathcal{U}\) are invariant under this reparametrization, while the remaining functions transform covariantly. We will discuss this remaining freedom in section~\ref{sec:multi}.

\subsection{Einstein-like frame}\label{ssec:einstein}
Another commonly used frame in the scalar-curvature class of gravity theories is the Einstein frame, in which there is no direct coupling between the scalar field and the Ricci scalar in the gravitational part of the action~\cite{Flanagan:2004bz}. A similar frame has been found also in a class of scalar-torsion theories which is invariant under conformal rescalings of the tetrad~\cite{Hohmann:2018ijr}. For the class of general scalar-teleparallel theories which we study in this article, one could similarly impose the condition \(\mathcal{A} \equiv 1\). However, note that this would determine only the function \(\gamma\) defining the conformal transformation of the metric, while the action~\eqref{eq:genaction} does not offer any preferred choice for the function \(\zeta\) defining the transformation of the teleparallel affine connection. This is different for the scalar-nonmetricity and scalar-torsion classes presented in sections~\ref{ssec:symacttra} and~\ref{ssec:metacttra}, in which there exists another function \(\mathcal{T}\) or \(\mathcal{Q}\), which defines the algebraic constraint on the torsion or nonmetricity, respectively. One can thus define a frame by imposing the additional condition that the corresponding function vanishes, so that the connection becomes either symmetric or metric-compatible, which then fixes the function \(\zeta\). We will discuss these definitions of an Einstein frame for the two relevant classes of theories below.

We start with the generalized class of scalar-nonmetricity theories defined in section~\ref{ssec:symacttra}. In this case we define the Einstein frame by imposing vanishing coupling of the scalar field to the nonmetricity scalar and vanishing torsion of the connection, which translates to the conditions
\begin{equation}\label{eq:efsymcond}
\ef{\st{\mathcal{A}}} \equiv 1\,, \quad
\ef{\mathcal{T}} \equiv 0\,.
\end{equation}
By comparison to the transformation rules~\eqref{eq:symcontra} and~\eqref{eq:symfuntra}, we see that this frame is obtained from an arbitrary frame by applying the transformation defined by
\begin{equation}\label{eq:efsymtrans}
\ef{\gamma} = \frac{1}{2}\ln\st{\mathcal{A}}\,, \quad
\ef{\zeta} = \mathcal{T}\,,
\end{equation}
under which the metric and the connection become
\begin{equation}
\ef{g}_{\mu\nu} = \st{\mathcal{A}}g_{\mu\nu}\,, \quad
\ef{\Gamma}^{\mu}{}_{\nu\rho} = \Gamma^{\mu}{}_{\nu\rho} + \mathcal{T}\delta^{\mu}_{\nu}\phi_{,\rho}\,.
\end{equation}
As for the Jordan frame, these are invariant field variables, in the sense that their definition does not depend on the choice of the original frame. Finally, we are left with expressing the remaining functions in the action, which now includes also the functions \(\alpha\) and \(\beta\) defining the matter coupling of the scalar field, in terms of invariant quantities. In this case we find that they are given by
\begin{equation}
\ef{\st{\mathcal{B}}} = 2\st{\mathcal{F}}\,, \quad
\ef{\st{\mathcal{D}}} = 2\st{\mathcal{J}}\,, \quad
\ef{\st{\mathcal{E}}} = 2\st{\mathcal{L}}\,, \quad
\ef{\st{\mathcal{V}}} = \mathcal{U}\,, \quad
\ef{\alpha} = \frac{1}{2}\ln\mathcal{I}\,, \quad
\ef{\beta} = -\mathcal{S}\,.
\end{equation}
By comparison with their values~\eqref{eq:jfsymfunc} in the Jordan frame, we now see that the invariants \(\mathcal{I}\) and \(\mathcal{S}\), which previously defined the coupling to the nonmetricity scalar and the torsion constraint, now define the matter coupling.

We then continue with the generalized scalar-torsion class of theories. In this case we define the Einstein frame by imposing vanishing coupling to the torsion scalar and vanishing nonmetricity, from which we obtain the conditions
\begin{equation}\label{eq:efmetcond}
\ef{\mt{\mathcal{A}}} \equiv 1\,, \quad
\ef{\mathcal{Q}} \equiv 0
\end{equation}
on the parameter functions in the gravitational part of the action. From the transformation rules~\eqref{eq:metcontra} and~\eqref{eq:metfuntra} we then find that this frame is related to an arbitrary frame by the transformation
\begin{equation}\label{eq:efmettrans}
\ef{\gamma} = \frac{1}{2}\ln\mt{\mathcal{A}}\,, \quad
\ef{\zeta} = \mathcal{Q} + \frac{\mt{\mathcal{A}}'}{2\mt{\mathcal{A}}}\,.
\end{equation}
In this case the invariant metric and connection thus read
\begin{equation}
\ef{g}_{\mu\nu} = \mt{\mathcal{A}}g_{\mu\nu}\,, \quad
\ef{\Gamma}^{\mu}{}_{\nu\rho} = \Gamma^{\mu}{}_{\nu\rho} + \left(\mathcal{Q} + \frac{\mt{\mathcal{A}}'}{2\mt{\mathcal{A}}}\right)\delta^{\mu}_{\nu}\phi_{,\rho}\,.
\end{equation}
Finally, the parameter functions in the action are now expressed in terms of invariant quantities as
\begin{equation}
\ef{\mt{\mathcal{B}}} = 2\mt{\mathcal{F}}\,, \quad
\ef{\mt{\mathcal{C}}} = 2\mt{\mathcal{H}}\,, \quad
\ef{\mt{\mathcal{V}}} = \mathcal{U}\,, \quad
\ef{\alpha} = \frac{1}{2}\ln\mathcal{I}\,, \quad
\ef{\beta} = \frac{\mathcal{I}'}{2\mathcal{I}} - \mathcal{P}\,.
\end{equation}
As in the scalar-nonmetricity case, we see that the invariants \(\mathcal{I}\) and \(\mathcal{P}\), which define the coupling to the torsion scalar and the nonmetricity through the relations~\eqref{eq:jfmetfunc} in the Jordan frame, now define the matter coupling. We will return to this observation in section~\ref{sec:char}, when we give a physical interpretation to these invariants.

\section{Multi-scalar-teleparallel extension}\label{sec:multi}
So far we have considered only scalar-teleparallel gravity theories in which there exists a single scalar field as a dynamical field variable next to the metric and flat affine connection. We now show how our results are generalized to the case of multiple scalar fields, hence giving rise to multi-scalar-teleparallel theories of gravity. In this section we thus replace the single scalar field \(\phi\) by a scalar field multiplet \(\boldsymbol{\phi} = (\phi^a, a = 1, \ldots, N)\) of \(N\) scalar fields. This is motivated by a similar generalization which can be constructed in scalar-curvature~\cite{Damour:1992we,Kuusk:2015dda} and scalar-torsion~\cite{Hohmann:2018ijr} theories of gravity. For the scalar-teleparallel we study in this article, a similar generalization is possible, but not without intricacies, which we discuss in this section. We start with a discussion of the field transformations in section~\ref{ssec:mfieldtra}. The multi-scalar-teleparallel action functionals are defined in section~\ref{ssec:maction}, and their transformation is shown in section~\ref{ssec:mactiontrans}. From these transformations, we obtain invariant quantities in section~\ref{ssec:minvariant}. Finally, we generalize the notion of Jordan and Einstein frames to multi-scalar-teleparallel theories in section~\ref{ssec:mframes}.

\subsection{Field transformation}\label{ssec:mfieldtra}
We start our discussion of the multi-scalar-teleparallel case by defining the transformation of the fundamental fields and studying the resulting transformation of derived geometric objects. For the scalar fields, this means that we have to replace the single function \(f\) by \(N\) functions \(f^a\), which define the transformation
\begin{equation}\label{eq:mscaltrans}
\bar{\phi}^a = f^a(\boldsymbol{\phi})\,,
\end{equation}
and now depend on all scalar fields in the multiplet \(\boldsymbol{\phi}\). As in the single-field case, we restrict ourselves to functions \(f^a\) such that this transformation is invertible. For the metric, the transformation reads
\begin{equation}\label{eq:mmettrans}
\bar{g}_{\mu\nu} = g_{\mu\nu}e^{2\gamma(\boldsymbol{\phi})}\,,
\end{equation}
and thus the only change is the fact that now also the function \(\gamma\) depends on all scalar fields. Finally, for the connection we study a generalized transformation of the form
\begin{equation}\label{eq:mconntrans}
\bar{\Gamma}^{\mu}{}_{\nu\rho} = \Gamma^{\mu}{}_{\nu\rho} + \zeta_a(\boldsymbol{\phi})\delta^{\mu}_{\nu}\phi^a_{,\rho}
\end{equation}
defined by \(N\) functions \(\zeta_a\), where we have chosen the index position such that we can make use of the Einstein summation convention also for indices which label scalar fields, as it will turn out to be well-defined throughout the remainder of this article. We can then study the transformation of derived geometric objects. For the torsion, we find the transformation
\begin{equation}\label{eq:mtortrans}
\bar{T}^{\mu}{}_{\nu\rho} = T^{\mu}{}_{\nu\rho} - 2\zeta_a\delta^{\mu}_{[\nu}\phi^a_{,\rho]}\,,
\end{equation}
while the transformation of the nonmetricity reads
\begin{equation}\label{eq:mnomtrans}
\bar{Q}_{\mu\nu\rho} = e^{2\gamma}[Q_{\mu\nu\rho} - 2(\zeta_a - \gamma_{,a})g_{\nu\rho}\phi^a_{,\mu}]\,.
\end{equation}
Here and in the following, a subscript with a comma denotes a derivative of a function with respect to the corresponding scalar field. For the curvature, however, we now find a non-vanishing contribution
\begin{equation}\label{eq:mcurtrans}
\bar{R}^{\mu}{}_{\nu\rho\sigma} = R^{\mu}{}_{\nu\rho\sigma} - 2 \zeta_{[a,b]}\delta^{\mu}_{\nu}\phi^a_{,\rho}\phi^b_{,\sigma}\,.
\end{equation}
Note the appearance of a term \(\zeta_{[a,b]}\), which vanishes in the single-field case. One obvious possibility to maintain vanishing curvature is to consider only transformations of the form \(\zeta_a = \xi_{,a}\) with some function \(\xi(\boldsymbol{\phi})\) of the scalar fields, such that this condition is automatically satisfied. This is motivated by the geometric interpretation given in section~\ref{sec:geom} that locally a flat connection corresponds to a path-independent parallel transport described by a locally transported basis \(\Lambda^{\mu}{}_{\nu}\), which we can now transform with a scalar-field dependent rescaling defined by
\begin{equation}
\bar{\Lambda}^{\mu}{}_{\nu} = \xi(\boldsymbol{\phi})\Lambda^{\mu}{}_{\nu}\,,
\end{equation}
such that the connection defined by \(\bar{\Lambda}^{\mu}{}_{\nu}\) has the coefficients
\begin{equation}
\bar{\Gamma}^{\mu}{}_{\nu\rho} = \Gamma^{\mu}{}_{\nu\rho} + \xi_{,a}(\boldsymbol{\phi})\delta^{\mu}_{\nu}\phi^a_{,\rho}\,.
\end{equation}
Alternatively, one may proceed as for the torsion and nonmetricity before and enlarge the class of theories and also study theories in which the curvature is non-vanishing, but determined by an algebraic constraint equation. While the latter would exceed the scope of this article, we will keep \(\zeta_a\) arbitrary wherever possible, and remark on the implications and necessity of setting \(\zeta_a = \xi_{,a}\) wherever it leads to any non-trivial consequences.

\subsection{Scalar-teleparallel action}\label{ssec:maction}
Before we can study the transformation of multi-scalar-teleparallel gravity theories, we first need to define their action functionals, which now also involve all scalar fields. We start with the matter action, which we now assume to be of the generalized form
\begin{equation}\label{eq:mmataction}
S_{\text{m}}\left[g_{\mu\nu}, \Gamma^{\mu}{}_{\nu\rho}, \phi^a, \chi^I\right] = \hat{S}_{\text{m}}\left[g_{\mu\nu}e^{2\alpha(\boldsymbol{\phi})}, \Gamma^{\mu}{}_{\nu\rho} + \beta_a(\boldsymbol{\phi})\delta^{\mu}_{\nu}\phi^a_{,\rho}, \chi^I\right]\,,
\end{equation}
with a single function \(\alpha\) and \(N\) functions \(\beta_a\) of the scalar field. For the variation of the matter action we write
\begin{equation}\label{eq:mmatactvar}
\delta S_{\text{m}} = \int_M\left(\frac{1}{2}\Theta^{\mu\nu}\delta g_{\mu\nu} + H_{\mu}{}^{\nu\rho}\delta\Gamma^{\mu}{}_{\nu\rho} + \Phi_a\delta\phi^a + \varpi_I\delta\chi^I\right)\sqrt{-g}\dd^4x\,,
\end{equation}
where we now have \(N\) terms \(\Phi_a\). As in the single-field case, it follows from the structure~\eqref{eq:mmataction} of the action that these terms \(\Phi_a\) are not independent, but can be expressed through the energy-momentum and hypermomentum tensors as
\begin{equation}\label{eq:mmatscal}
\Phi_a = \alpha_{,a}\Theta - \beta_a\lc{\nabla}_{\nu}H_{\mu}{}^{\mu\nu}\,.
\end{equation}
We then turn our focus to the gravitational part of the action. First, note that the scalar terms in the action which involve the scalar field generalize to
\begin{equation}
X^{ab} = -\frac{1}{2}g^{\mu\nu}\phi^a_{,\mu}\phi^b_{,\nu}\,, \quad
U^a = T_{\mu}{}^{\mu\nu}\phi^a_{,\nu}\,, \quad
V^a = Q^{\nu\mu}{}_{\mu}\phi^a_{,\nu}\,, \quad
W^a = Q_{\mu}{}^{\mu\nu}\phi^a_{,\nu}\,,
\end{equation}
where \(X^{ab}\) is obviously symmetric in its indices. This larger number of scalar quantities, which are now organized in multiplets, must also be reflected by the parameter functions, which govern their contribution to the gravitational action. For the general scalar-teleparallel action this means that it now becomes
\begin{equation}\label{eq:mgenaction}
S_G[g_{\mu\nu}, \Gamma^{\mu}{}_{\nu\rho}, \phi^a] = \frac{1}{2\kappa^2}\int_M\left[-\mathcal{A}(\boldsymbol{\phi})G + 2\mathcal{B}_{ab}(\boldsymbol{\phi})X^{ab} + 2\mathcal{C}_a(\boldsymbol{\phi})U^a + 2\mathcal{D}_a(\boldsymbol{\phi})V^a + 2\mathcal{E}_a(\boldsymbol{\phi})W^a - 2\kappa^2\mathcal{V}(\boldsymbol{\phi})\right]\sqrt{-g}\dd^4x\,,
\end{equation}
and so it depends on functions \(\mathcal{A}, \mathcal{B}_{ab}, \mathcal{C}_a, \mathcal{D}_a, \mathcal{E}_a, \mathcal{V}\). In order to proceed towards the symmetric and teleparallel cases, we need to consider the generalized Lagrange multiplier terms \(S_{\mathfrak{t}}'\) and \(S_{\mathfrak{q}}'\) introduced in section~\ref{sec:actiontrans}. A naive generalization of these terms to multiple scalar fields reads
\begin{equation}\label{eq:mtorlagmul}
S_{\mathfrak{t}}'[\mathfrak{t}_{\mu}{}^{\nu\rho}, \Gamma^{\mu}{}_{\nu\rho}, \phi^a] = \int_M\mathfrak{t}_{\mu}{}^{\nu\rho}\left[T^{\mu}{}_{\nu\rho} - 2\mathcal{T}_a(\boldsymbol{\phi})\delta^{\mu}_{[\nu}\phi^a_{,\rho]}\right]\dd^4x
\end{equation}
and
\begin{equation}\label{eq:mnomlagmul}
S_{\mathfrak{q}}'[\mathfrak{q}^{\mu\nu\rho}, g_{\mu\nu}, \Gamma^{\mu}{}_{\nu\rho}, \phi^a] = \int_M\mathfrak{q}^{\mu\nu\rho}\left[Q_{\mu\nu\rho} - 2\mathcal{Q}_a(\boldsymbol{\phi})g_{\nu\rho}\phi^a_{,\mu}\right]\dd^4x\,,
\end{equation}
which now each depend on \(N\) functions \(\mathcal{T}_a\) and \(\mathcal{Q}_a\), respectively. However, note that constraining the torsion and the nonmetricity also imposes a constraint on the curvature through the Bianchi identities
\begin{equation}
R^{\mu}{}_{[\nu\rho\sigma]} = \nabla_{[\nu}T^{\mu}{}_{\rho\sigma]} + T^{\mu}{}_{\omega[\nu}T^{\omega}{}_{\rho\sigma]} = 2\delta^{\mu}_{[\nu}\phi^a_{,\rho}\phi^b_{,\sigma]}\mathcal{T}_{a,b}
\end{equation}
and
\begin{equation}
2R_{(\rho\sigma)\mu\nu} = 2\nabla_{[\mu}Q_{\nu]\rho\sigma} + T^{\omega}{}_{\mu\nu}Q_{\omega\rho\sigma} = 4\phi^a_{,[\mu}\phi^b_{,\nu]}g_{\rho\sigma}\mathcal{Q}_{a,b}\,.
\end{equation}
We see that in order for these to vanish for arbitrary scalar field configurations we must impose the restrictions \(\mathcal{T}_{[a,b]} \equiv 0\) and \(\mathcal{Q}_{[a,b]} \equiv 0\), since otherwise these constraints would not be compatible with the flatness constraint. Imposing the constraint~\eqref{eq:mtorlagmul} on the torsion, we find that the terms in the multi-scalar-teleparallel action reduce to
\begin{equation}
G = Q + 2\mathcal{T}_a(V^a - W^a) + 12\mathcal{T}_a\mathcal{T}_bX^{ab}\,, \quad
U^a = -6\mathcal{T}_bX^{ab}\,,
\end{equation}
while the constraint~\eqref{eq:mnomlagmul} for the nonmetricity yields
\begin{equation}
G = T + 4\mathcal{Q}_aU^a + 12\mathcal{Q}_a\mathcal{Q}_bX^{ab}\,, \quad
V^a = -16\mathcal{Q}_bX^{ab}\,, \quad
W^a = -4\mathcal{Q}_bX^{ab}\,.
\end{equation}
It follows that we can write the multi-scalar-nonmetricity action as
\begin{equation}\label{eq:msymaction}
S_Q'[g_{\mu\nu}, \Gamma^{\mu}{}_{\nu\rho}, \phi^a] = \frac{1}{2\kappa^2}\int_M\left[-\st{\mathcal{A}}(\boldsymbol{\phi})Q + 2\st{\mathcal{B}}_{ab}(\boldsymbol{\phi})X^{ab} + 2\st{\mathcal{D}}_a(\boldsymbol{\phi})V^a + 2\st{\mathcal{E}}_a(\boldsymbol{\phi})W^a - 2\kappa^2\st{\mathcal{V}}(\boldsymbol{\phi})\right]\sqrt{-g}\dd^4x\,,
\end{equation}
while the multi-scalar-torsion action becomes
\begin{equation}\label{eq:mmetaction}
S_T'[g_{\mu\nu}, \Gamma^{\mu}{}_{\nu\rho}, \phi^a] = \frac{1}{2\kappa^2}\int_M\left[-\mt{\mathcal{A}}(\boldsymbol{\phi})T + 2\mt{\mathcal{B}}_{ab}(\boldsymbol{\phi})X^{ab} + 2\mt{\mathcal{C}}_a(\boldsymbol{\phi})U^a - 2\kappa^2\mt{\mathcal{V}}(\boldsymbol{\phi})\right]\sqrt{-g}\dd^4x\,,
\end{equation}
where the appearing parameter functions are related by
\begin{equation}\label{eq:msymactfunc}
\st{\mathcal{A}} = \mathcal{A}\,, \quad
\st{\mathcal{B}}_{ab} = \mathcal{B}_{ab} - 6\mathcal{C}_{(a}\mathcal{T}_{b)} - 6\mathcal{A}\mathcal{T}_a\mathcal{T}_b\,, \quad
\st{\mathcal{D}}_a = \mathcal{D}_a - \mathcal{A}\mathcal{T}_a\,, \quad
\st{\mathcal{E}}_a = \mathcal{E}_a + \mathcal{A}\mathcal{T}_a\,, \quad
\st{\mathcal{V}} = \mathcal{V}
\end{equation}
and
\begin{equation}\label{eq:mmetactfunc}
\mt{\mathcal{A}} = \mathcal{A}\,, \quad
\mt{\mathcal{B}}_{ab} = \mathcal{B}_{ab} - 16\mathcal{D}_{(a}\mathcal{Q}_{b)} - 4\mathcal{E}_{(a}\mathcal{Q}_{b)} - 6\mathcal{A}\mathcal{Q}_a\mathcal{Q}_b\,, \quad
\mt{\mathcal{C}}_a = \mathcal{C}_a - 2\mathcal{A}\mathcal{Q}_a\,, \quad
\mt{\mathcal{V}} = \mathcal{V}\,,
\end{equation}
respectively. This completes the family of multi-scalar-teleparallel gravity actions, which we discuss in the following.

\subsection{Transformation of multi-scalar-teleparallel gravity}\label{ssec:mactiontrans}
We now study the transformation of the various multi-scalar-teleparallel gravity actions introduced in section~\ref{ssec:maction} under the generalized field transformations introduced in section~\ref{ssec:mfieldtra}, following the procedure detailed for the single-field case in section~\ref{sec:actiontrans}. We start with the matter action~\eqref{eq:mmataction}. Following the same steps as outlined for the single-field case in section~\ref{ssec:matacttra}, we find that the functions defining the matter coupling transform as
\begin{equation}\label{eq:mmatfuntra}
\alpha = \bar{\alpha} + \gamma\,, \quad
\beta_a = f^b_{,a}\bar{\beta}_b + \zeta_a\,.
\end{equation}
We then turn our attention towards the gravitational part of the action. First note that the scalar quantities constructed from the field variables undergo the transformations
\begin{subequations}\label{eq:mgrascaltra}
\begin{align}
\bar{G} &= e^{-2\gamma}[G + 2\gamma_{,a}(2U^a - V^a + W^a) + 12\gamma_{,a}\gamma_{,b}X^{ab}]\,,\\
\bar{T} &= e^{-2\gamma}(T + 4\zeta_aU^a + 12\zeta_a\zeta_bX^{ab})\,,\\
\bar{Q} &= e^{-2\gamma}[Q + 2(\zeta_a - \gamma_{,a})(V^a - W^a) + 12(\zeta_a - \gamma_{,a})(\zeta_b - \gamma_{,b})X^{ab}]\,,\\
\bar{X}^{ab} &= e^{-2\gamma}f^a_{,c}f^b_{,d}X^{cd}\,,\\
\bar{U}^a &= e^{-2\gamma}f^a_{,b}(U^b + 6\zeta_cX^{bc})\,,\\
\bar{V}^a &= e^{-2\gamma}f^a_{,b}[V^b + 16(\zeta_c - \gamma_{,c})X^{bc}]\,,\\
\bar{W}^a &= e^{-2\gamma}f^a_{,b}[W^b + 4(\zeta_c - \gamma_{,c})X^{bc}]\,.
\end{align}
\end{subequations}
With the help of these formulas, we can study the transformation of the different contributions to the gravitational part of the action. From the general multi-scalar-teleparallel action~\eqref{eq:mgenaction}, we read off the transformations
\begin{subequations}\label{eq:mgenfuntra}
\begin{align}
\mathcal{A} &= e^{2\gamma}\bar{\mathcal{A}}\,,\\
\mathcal{B}_{ab} &= e^{2\gamma}[f^c_{,a}f^d_{,b}\bar{\mathcal{B}}_{cd} - 6\gamma_{,a}\gamma_{,b}\bar{\mathcal{A}} + 6\zeta_{(a}f^c_{,b)}\bar{\mathcal{C}}_c + 4(\zeta_{(a} - \gamma_{(a,})f^c_{,b)}(4\bar{\mathcal{D}}_c + \bar{\mathcal{E}}_c)]\,,\\
\mathcal{C}_a &= e^{2\gamma}(f^b_{,a}\bar{\mathcal{C}}_b - 2\gamma_{,a}\bar{\mathcal{A}})\,,\\
\mathcal{D}_a &= e^{2\gamma}(f^b_{,a}\bar{\mathcal{D}}_b + \gamma_{,a}\bar{\mathcal{A}})\,,\\
\mathcal{E}_a &= e^{2\gamma}(f^b_{,a}\bar{\mathcal{E}}_b - \gamma_{,a}\bar{\mathcal{A}})\,,\\
\mathcal{V} &= e^{4\gamma}\bar{\mathcal{V}}\,.
\end{align}
\end{subequations}
Similarly, for the multi-scalar-nonmetricity action~\eqref{eq:msymaction} we find the transformations
\begin{subequations}\label{eq:msymfuntra}
\begin{align}
\st{\mathcal{A}} &= e^{2\gamma}\st{\bar{\mathcal{A}}}\,,\\
\st{\mathcal{B}}_{ab} &= e^{2\gamma}\left[f^c_{,a}f^d_{,b}\st{\bar{\mathcal{B}}}_{cd} - 6(\zeta_a - \gamma_{,a})(\zeta_b - \gamma_{,b})\st{\bar{\mathcal{A}}} + 4(\zeta_{(a} - \gamma_{,(a})f^c_{,b)}\left(4\st{\bar{\mathcal{D}}}_c + \st{\bar{\mathcal{E}}}_c\right)\right]\,,\\
\st{\mathcal{D}}_a &= e^{2\gamma}\left[f^b_{,a}\st{\bar{\mathcal{D}}}_{,b} - (\zeta_a - \gamma_{,a})\st{\bar{\mathcal{A}}}\right]\,,\\
\st{\mathcal{E}}_a &= e^{2\gamma}\left[f^b_{,a}\st{\bar{\mathcal{E}}}_{,b} + (\zeta_a - \gamma_{,a})\st{\bar{\mathcal{A}}}\right]\,,\\
\st{\mathcal{V}} &= e^{4\gamma}\st{\bar{\mathcal{V}}}\,,
\end{align}
\end{subequations}
while the multi-scalar-torsion action~\eqref{eq:mmetaction} transforms as
\begin{subequations}\label{eq:mmetfuntra}
\begin{align}
\mt{\mathcal{A}} &= e^{2\gamma}\mt{\bar{\mathcal{A}}}\,,\\
\mt{\mathcal{B}}_{ab} &= e^{2\gamma}\left(f^c_{,a}f^d_{,b}\mt{\bar{\mathcal{B}}}_{cd} - 6\zeta_a\zeta_b\mt{\bar{\mathcal{A}}} + 6\zeta_{(a}f^c_{,b)}\mt{\bar{\mathcal{C}}}_c\right)\,,\\
\mt{\mathcal{C}}_a &= e^{2\gamma}\left(f^b_{,a}\mt{\bar{\mathcal{C}}}_b - 2\zeta_a\mt{\bar{\mathcal{A}}}\right)\,,\\
\mt{\mathcal{V}} &= e^{4\gamma}\mt{\bar{\mathcal{V}}}\,.
\end{align}
\end{subequations}
To complete our discussion, we also need to consider the Lagrange multiplier terms, which enforce the vanishing curvature, torsion and nonmetricity, respectively. From the latter two, given by the functionals~\eqref{eq:mtorlagmul} and~\eqref{eq:mnomlagmul}, we find the transformations
\begin{equation}
\mathcal{T}_a = f^b_{,a}\bar{\mathcal{T}}_b + \zeta_a\,, \quad
\mathcal{Q}_a = f^b_{,a}\bar{\mathcal{Q}}_b + \zeta_a - \gamma_{,a}\,.
\end{equation}
Finally, we must also consider the term~\eqref{eq:curlagmul}, whose transformation now becomes
\begin{equation}\label{eq:mcurlagtra}
\bar{S}_{\mathfrak{r}}[\bar{\mathfrak{r}}_{\mu}{}^{\nu\rho\sigma}, \bar{\Gamma}^{\mu}{}_{\nu\rho}] = \int_M\bar{\mathfrak{r}}_{\mu}{}^{\nu\rho\sigma}\bar{R}^{\mu}{}_{\nu\rho\sigma}\dd^4x = \int_M\bar{\mathfrak{r}}_{\mu}{}^{\nu\rho\sigma}(R^{\mu}{}_{\nu\rho\sigma} - 2\delta^{\mu}_{\nu}\phi^a_{,\rho}\phi^b_{\sigma}\zeta_{[a,b]})\dd^4x\,,
\end{equation}
and so in contrast to the single-field case it does not retain its form, unless \(\zeta_{[a,b]} \equiv 0\). One possibility would be to proceed in analogy to the Lagrange multiplier terms for the torsion and nonmetricity, and consider a more general term
\begin{equation}
S'_{\mathfrak{r}}[\mathfrak{r}_{\mu}{}^{\nu\rho\sigma}, \Gamma^{\mu}{}_{\nu\rho}, \phi] = \int_M\mathfrak{r}_{\mu}{}^{\nu\rho\sigma}\left[R^{\mu}{}_{\nu\rho\sigma} - 2\delta^{\mu}_{\nu}\phi^a_{,\rho}\phi^b_{\sigma}\mathcal{R}_{[a,b]}\right]\dd^4x\,,
\end{equation}
which includes new functions \(\mathcal{R}_a\) of the scalar field, and imposes the constraint
\begin{equation}
R^{\mu}{}_{\nu\rho\sigma} = 2\delta^{\mu}_{\nu}\phi^a_{,\rho}\phi^b_{\sigma}\mathcal{R}_{[a,b]}
\end{equation}
on the curvature of the connection. Similar considerations would also be necessary if we abandoned the restrictions \(\mathcal{T}_{[a,b]} \equiv 0\) or \(\mathcal{Q}_{[a,b]} \equiv 0\). However, this would lead to a non-flat connection, which has a number of consequences, which must be taken into account. Most notably, the difference between the scalar \(\mathcal{G}\) and the Ricci scalar \(\lc{R}\) is no longer a boundary term, which invalidates the GR limit of the theory given by the GTEGR action, and so the action \(S_G\) must be modified accordingly to include also curvature terms. This would extend the discussion into the general metric-affine class of theories, which exceeds the scope of this article; see~\cite{Iosifidis:2018zwo} for possible extensions. Here we restrict ourselves to the original flatness constraint~\eqref{eq:curlagmul}, and thus consider only such transformations for which \(\zeta_a \equiv \xi_{,a}\) for some function \(\xi\) of the scalar fields, such that \(\zeta_{[a,b]} \equiv 0\).

\subsection{Invariant quantities}\label{ssec:minvariant}
Using the transformation behavior of the parameter functions in the multi-scalar-teleparallel gravity action, it is now straightforward to generalize the construction of invariant quantities constructed in the single-field case. First, note that for the functions \(\mathcal{A}\), \(\mathcal{V}\) and \(\alpha\) there is no significant difference to the single-field case, and so the invariants \(\mathcal{I}\) and \(\mathcal{U}\) are defined in full analogy. Note that these functions now depend on all scalar fields, and so in place of their (single) derivative, which we used to construct further invariants, we now have the derivatives
\begin{equation}
\mathcal{A}_{,a} = e^{2\gamma}(2\gamma_{,a}\bar{\mathcal{A}} + f^b_{,a}\bar{\mathcal{A}}_{,b})\,, \quad
\alpha_{,a} = f^b_{,a}\bar{\alpha}_{,b} + \gamma_{,a}\,.
\end{equation}
Together with the functions \(\mathcal{C}_a\), \(\mathcal{D}_a\) and \(\mathcal{E}_a\), we can use these to construct the invariants
\begin{equation}\label{eq:mvecinv1}
\mathcal{K}_a = \frac{\mathcal{C}_a + 2\alpha_{,a}\mathcal{A}}{2e^{2\alpha}}\,, \quad
\mathcal{M}_a = \frac{\mathcal{D}_a - \alpha_{,a}\mathcal{A}}{2e^{2\alpha}}\,, \quad
\mathcal{N}_a = \frac{\mathcal{E}_a + \alpha_{,a}\mathcal{A}}{2e^{2\alpha}}\,,
\end{equation}
and similarly
\begin{equation}\label{eq:mvecinv2}
\mathcal{H}_a = \frac{\mathcal{C}_a + \mathcal{A}_{,a}}{2\mathcal{A}}\,, \quad
\mathcal{J}_a = \frac{2\mathcal{D}_a - \mathcal{A}_{,a}}{4\mathcal{A}}\,, \quad
\mathcal{L}_a = \frac{2\mathcal{E}_a + \mathcal{A}_{,a}}{4\mathcal{A}}\,.
\end{equation}
As in the single-field case, these are invariant only under transformations of the metric and the connection with functions \(\gamma\) and \(\zeta_a\), while under scalar field transformations they transform covariantly as
\begin{equation}\label{eq:mcovtra1}
\mathcal{K}_a = f^b_{,a}\bar{\mathcal{K}}_b\,,
\end{equation}
and analogously for the other invariants. The same holds also for the quantities
\begin{equation}
\mathcal{S}_a = \mathcal{T}_a - \beta_a\,, \quad
\mathcal{P}_a = \mathcal{Q}_a + \alpha_{,a} - \beta_a\,,
\end{equation}
which are constructed from \(\mathcal{T}_a\), \(\mathcal{Q}_a\) and \(\beta_a\). Finally, the term constructed from \(\mathcal{B}_{ab}\) now takes the form
\begin{equation}
\mathcal{G}_{ab} = \frac{\mathcal{B}_{ab} - 6\alpha_{,a}\alpha_{,b}\mathcal{A} - 6\beta_{(a}\mathcal{C}_{b)} + 4(\alpha_{,(a} - \beta_{(a})(4\mathcal{D}_{b)} + \mathcal{E}_{b)})}{2e^{2\alpha}}\,,
\end{equation}
where we must take care of the symmetry in its two indices, and transforms as
\begin{equation}\label{eq:mcovtra2}
\mathcal{G}_{ab} = f^c_{,a}f^d_{,b}\bar{\mathcal{G}}_{cd}\,.
\end{equation}
By comparison of the transformation rules~\eqref{eq:mcovtra1} and~\eqref{eq:mcovtra2}, which generalize the rules~\eqref{eq:covtra1} and~\eqref{eq:covtra2} found in the single-field case, we now see that \(\mathcal{G}_{ab}\) transforms as the components of a symmetric rank-two tensor on the space of scalar fields, while the previously introduced invariants which carry one index transform as the components of a covector. Keeping this geometric interpretation in mind, it is not surprising that the (exterior) derivative of a scalar invariant such as \(\mathcal{I}\) yields the components of a vector invariant. This becomes apparent, e.g., in the relations
\begin{equation}
2(\mathcal{I}\mathcal{K}_a - \mathcal{H}_a) = -4(\mathcal{I}\mathcal{M}_a - \mathcal{J}_a) = 4(\mathcal{I}\mathcal{N}_a - \mathcal{L}_a) = \frac{\mathcal{I}_{,a}}{\mathcal{I}}\,.
\end{equation}
Also in the multi-scalar-teleparallel case we can proceed to construct invariants also for the symmetric and metric classes of theories. For the vector invariants, one finds the straightforward generalizations
\begin{equation}\label{eq:mvecinv3}
\mt{\mathcal{K}}_a = \frac{\mt{\mathcal{C}}_a + 2\beta\mt{\mathcal{A}}}{2e^{2\alpha}}\,, \quad
\st{\mathcal{M}}_a = \frac{\st{\mathcal{D}}_a - (\alpha' - \beta)\st{\mathcal{A}}}{2e^{2\alpha}}\,, \quad
\st{\mathcal{N}}_a = \frac{\st{\mathcal{E}}_a + (\alpha' - \beta)\st{\mathcal{A}}}{2e^{2\alpha}}
\end{equation}
and
\begin{equation}\label{eq:mvecinv4}
\mt{\mathcal{H}}_a = \frac{\mt{\mathcal{C}}_a + \mt{\mathcal{A}}_{,a} + 2\mt{\mathcal{A}}\mathcal{Q}_a}{2\mt{\mathcal{A}}}\,, \quad
\st{\mathcal{J}}_a = \frac{2\st{\mathcal{D}}_a - \st{\mathcal{A}}_{,a} + 2\st{\mathcal{A}}\mathcal{T}_a}{4\st{\mathcal{A}}}\,, \quad
\st{\mathcal{L}}_a = \frac{2\st{\mathcal{E}}_a + \st{\mathcal{A}}_{,a} - 2\st{\mathcal{A}}\mathcal{T}_a}{4\st{\mathcal{A}}}\,,
\end{equation}
which are expressed in terms of the previously constructed invariants as
\begin{equation}
\mt{\mathcal{K}}_a = \mathcal{K}_a - \frac{\mathcal{P}_a}{\mathcal{I}}\,,\quad
\st{\mathcal{M}}_a = \mathcal{M}_a - \frac{\mathcal{S}_a}{2\mathcal{I}}\,,\quad
\st{\mathcal{N}}_a = \mathcal{N}_a + \frac{\mathcal{S}_a}{2\mathcal{I}}\,,\quad
\mt{\mathcal{H}}_a = \mathcal{H}_a\,,\quad
\st{\mathcal{J}}_a = \mathcal{J}_a\,,\quad
\st{\mathcal{L}}_a = \mathcal{L}_a\,,
\end{equation}
and satisfy the relations
\begin{equation}
2\left(\mathcal{I}\mt{\mathcal{K}}_a - \mt{\mathcal{H}}_a\right) = \frac{\mathcal{I}_{,a}}{\mathcal{I}} - 2\mathcal{P}_a\,, \quad
-4\left(\mathcal{I}\st{\mathcal{M}}_a - \st{\mathcal{J}}_a\right) = 4\left(\mathcal{I}\st{\mathcal{N}}_a - \st{\mathcal{L}}_a\right) = \frac{\mathcal{I}_{,a}}{\mathcal{I}} + 2\mathcal{S}_a\,.
\end{equation}
Similarly, for the tensor invariants one finds the generalizations
\begin{equation}
\st{\mathcal{G}}_{ab} = \frac{\st{\mathcal{B}}_{ab} - 6(\alpha_{,a} - \beta_a)(\alpha_{,b} - \beta_b)\st{\mathcal{A}} + 4(\alpha_{,(a} - \beta_{(a})(4\st{\mathcal{D}}_{b)} + \st{\mathcal{E}}_{b)})}{2e^{2\alpha}}\,, \quad
\mt{\mathcal{G}}_{ab} = \frac{\mt{\mathcal{B}}_{ab} - 6\beta_a\beta_b\mt{\mathcal{A}} - 6\beta_{(a}\mt{\mathcal{C}}_{b)}}{2e^{2\alpha}}\,,
\end{equation}
as well as
\begin{subequations}
\begin{align}
\st{\mathcal{F}}_{ab} &= \frac{2\st{\mathcal{A}}\st{\mathcal{B}}_{ab} - 3(\st{\mathcal{A}}_{,a} - 2\st{\mathcal{A}}\mathcal{T}_a)(\st{\mathcal{A}}_{,b} - 2\st{\mathcal{A}}\mathcal{T}_b) + 4(\st{\mathcal{A}}_{,(a} - 2\st{\mathcal{A}}\mathcal{T}_{(a})(4\st{\mathcal{D}}_{b)} + \st{\mathcal{E}}_{b)})}{4\st{\mathcal{A}}^2}\,,\\
\mt{\mathcal{F}}_{ab} &= \frac{2\mt{\mathcal{A}}\mt{\mathcal{B}}_{ab} - 3(\mt{\mathcal{A}}_{,a} + 2\mt{\mathcal{A}}\mathcal{Q}_a)(\mt{\mathcal{A}}_{,b} + 2\mt{\mathcal{A}}\mathcal{Q}_b) - 6(\mt{\mathcal{A}}_{,(a} + 2\mt{\mathcal{A}}\mathcal{Q}_{(a})\mt{\mathcal{C}}_{b)}}{4\mt{\mathcal{A}}^2}\,.
\end{align}
\end{subequations}
In terms of the previously constructed invariants they read
\begin{equation}
\st{\mathcal{G}}_{ab} = \mathcal{G}_{ab} - 6\mathcal{K}_{(a}\mathcal{S}_{b)} - \frac{3\mathcal{S}_a\mathcal{S}_b}{\mathcal{I}}\,, \quad
\mt{\mathcal{G}}_{ab} = \mathcal{G}_{ab} - 4(4\mathcal{M}_{(a} + \mathcal{N}_{(a})\mathcal{P}_{b)} - \frac{3\mathcal{P}_a\mathcal{P}_b}{\mathcal{I}}
\end{equation}
and are related by
\begin{subequations}
\begin{align}
\st{\mathcal{F}}_{ab} &= \mathcal{I}\st{\mathcal{G}}_{ab} - 2(4\st{\mathcal{M}}_{(a} + \st{\mathcal{N}}_{(a})(2\mathcal{I}\mathcal{S}_{b)} + \mathcal{I}_{,b)}) - \frac{3(2\mathcal{I}\mathcal{S}_a + \mathcal{I}_{,a})(2\mathcal{I}\mathcal{S}_b + \mathcal{I}_{,b})}{4\mathcal{I}^2}\,,\\
\mt{\mathcal{F}}_{ab} &= \mathcal{I}\mt{\mathcal{G}}_{ab} - 3\mt{\mathcal{K}}_{(a}(2\mathcal{I}\mathcal{P}_{b)} - \mathcal{I}_{,b)}) - \frac{3(2\mathcal{I}\mathcal{P}_a - \mathcal{I}_{,a})(2\mathcal{I}\mathcal{P}_b - \mathcal{I}_{,b})}{4\mathcal{I}^2}\,.
\end{align}
\end{subequations}
This concludes our construction of invariants in the multi-scalar-teleparallel gravity theories we consider here, proceeding in full analogy to the single-field case. As in the latter, one may expect that the particular invariants shown here are of particular use in the formulation of such theories, and we will show this below. We finally remark that if we impose the restriction \(\zeta_{[a,b]} \equiv 0\) on the allowed transformations of the connection, in order to retain its flatness, then also quantities such as
\begin{equation}
\beta_{[a,b]}\,, \quad
\mathcal{T}_{[a,b]}\,, \quad
\mathcal{Q}_{[a,b]}
\end{equation}
are invariant under the restricted class of transformations, where we recall that the latter two must vanish as a consequence of the Bianchi identities in order to be compatible with vanishing curvature. We will see the relevance of such terms in the following section.

\subsection{Frames}\label{ssec:mframes}
In section~\ref{sec:frames} we have constructed scalar-teleparallel analogues to the Jordan and Einstein frames which are known from scalar-curvature gravity theories. We now generalize these frames to multi-scalar-teleparallel theories, where we must pay attention to possible restrictions arising from the condition \(\zeta_{[a,b]}\) which we impose to maintain vanishing curvature of the teleparallel affine connection. At first, we take a look at the Jordan frame, which we introduced in section~\ref{ssec:jordan}. If we simply generalize the condition~\eqref{eq:jfcond} to multi-scalar-teleparallel theories, it reads
\begin{equation}
\jf{\alpha} \equiv \jf{\beta}_a \equiv 0\,,
\end{equation}
and thus leads to the transformation
\begin{equation}
\jf{\gamma} = \alpha\,, \quad
\jf{\zeta}_a = \beta_a\,.
\end{equation}
However, keeping in mind the restriction \(\zeta_{[a,b]} \equiv 0\), we see that this is possible only if \(\beta_{[a,b]} \equiv 0\), unless one extends the class of theories beyond the teleparallel paradigm and also allows for algebraically determined, yet non-vanishing curvature. Since we do not delve into this extension in this article, we conclude that a true teleparallel Jordan frame exists in the multi-scalar-teleparallel case only for a particularly restricted class of matter couplings. Proceeding further in analogy to the single-field case, we then define the Jordan frame metric and connection by
\begin{equation}
\jf{g}_{\mu\nu} = e^{2\alpha}g_{\mu\nu}\,, \quad
\jf{\Gamma}^{\mu}{}_{\nu\rho} = \Gamma^{\mu}{}_{\nu\rho} + \beta_a\delta^{\mu}_{\nu}\phi^a_{,\rho}\,,
\end{equation}
and find that they are again invariant in the sense that
\begin{equation}
\jf{g}_{\mu\nu} = e^{2\alpha}g_{\mu\nu} = e^{2(\bar{\alpha} + \gamma)}g_{\mu\nu} = e^{2\bar{\alpha}}\bar{g}_{\mu\nu} = \jf{\bar{g}}_{\mu\nu}
\end{equation}
and
\begin{equation}
\jf{\Gamma}^{\mu}{}_{\nu\rho} = \Gamma^{\mu}{}_{\nu\rho} + \beta_a\delta^{\mu}_{\nu}\phi^a_{,\rho} = \Gamma^{\mu}{}_{\nu\rho} + (\bar{\beta}_a + \zeta_a)\delta^{\mu}_{\nu}\phi^a_{,\rho} = \bar{\Gamma}^{\mu}{}_{\nu\rho} + \bar{\beta}_a\delta^{\mu}_{\nu}\phi^a_{,\rho} = \jf{\bar{\Gamma}}^{\mu}{}_{\nu\rho}
\end{equation}
are defined independently of the choice of the original frame. Finally, for the parameter functions in the gravitational part of the action, we now find that in the general multi-scalar-teleparallel case they become
\begin{equation}
\jf{\mathcal{A}} = \frac{1}{\mathcal{I}}\,, \quad
\jf{\mathcal{B}}_{ab} = 2\mathcal{G}_{ab}\,, \quad
\jf{\mathcal{C}}_a = 2\mathcal{K}_a\,, \quad
\jf{\mathcal{D}}_a = 2\mathcal{M}_a\,, \quad
\jf{\mathcal{E}}_a = 2\mathcal{N}_a\,, \quad
\jf{\mathcal{V}} = \frac{\mathcal{U}}{\mathcal{I}^2}\,,
\end{equation}
while in the multi-scalar-nonmetricity theories we have
\begin{equation}
\jf{\st{\mathcal{A}}} = \frac{1}{\mathcal{I}}\,, \quad
\jf{\st{\mathcal{B}}}_{ab} = 2\st{\mathcal{G}}_{ab}\,, \quad
\jf{\st{\mathcal{D}}}_a = 2\st{\mathcal{M}}_a\,, \quad
\jf{\st{\mathcal{E}}}_a = 2\st{\mathcal{N}}_a\,, \quad
\jf{\st{\mathcal{V}}} = \frac{\mathcal{U}}{\mathcal{I}^2}\,, \quad
\jf{\mathcal{T}}_a = \mathcal{S}_a\,,
\end{equation}
and for the multi-scalar-torsion theories we find
\begin{equation}
\jf{\mt{\mathcal{A}}} = \frac{1}{\mathcal{I}}\,, \quad
\jf{\mt{\mathcal{B}}}_{ab} = 2\mt{\mathcal{G}}_{ab}\,, \quad
\jf{\mt{\mathcal{C}}}_a = 2\mt{\mathcal{K}}_a\,, \quad
\jf{\mt{\mathcal{V}}} = \frac{\mathcal{U}}{\mathcal{I}^2}\,, \quad
\jf{\mathcal{Q}}_a = \mathcal{P}_a\,,
\end{equation}
which is a straightforward generalization of the single-field case.

We then continue with the Einstein frame in generalized multi-scalar-nonmetricity theories, which we define by the conditions
\begin{equation}
\ef{\st{\mathcal{A}}} \equiv 1\,, \quad
\ef{\mathcal{T}}_a \equiv 0\,.
\end{equation}
Hence, the corresponding transformation from an arbitrary frame is given by
\begin{equation}
\ef{\gamma} = \frac{1}{2}\ln\st{\mathcal{A}}\,, \quad
\ef{\zeta}_a = \mathcal{T}_a\,,
\end{equation}
and also here we see that this is possible without violating the flatness constraint only for theories which satisfy \(\mathcal{T}_{[a,b]} \equiv 0\), as demanded by the Bianchi identities in this case. For theories which do not satisfy this condition it is not possible to find a frame in which both torsion and curvature vanish simultaneously, and so we see that such theories constitute a genuine extension to the purely multi-scalar-nonmetricity theories, which we do not study here. Assuming that the condition is satisfied, we can thus define the invariant metric and connection as
\begin{equation}
\ef{g}_{\mu\nu} = \st{\mathcal{A}}g_{\mu\nu}\,, \quad
\ef{\Gamma}^{\mu}{}_{\nu\rho} = \Gamma^{\mu}{}_{\nu\rho} + \mathcal{T}_a\delta^{\mu}_{\nu}\phi^a_{,\rho}\,,
\end{equation}
and finally express the multi-scalar-nonmetricity action through the invariants
\begin{equation}
\ef{\st{\mathcal{B}}}_{ab} = 2\st{\mathcal{F}}_{ab}\,, \quad
\ef{\st{\mathcal{D}}}_a = 2\st{\mathcal{J}}_a\,, \quad
\ef{\st{\mathcal{E}}}_a = 2\st{\mathcal{L}}_a\,, \quad
\ef{\st{\mathcal{V}}} = \mathcal{U}\,, \quad
\ef{\alpha} = \frac{1}{2}\ln\mathcal{I}\,, \quad
\ef{\beta}_a = -\mathcal{S}_a\,,
\end{equation}
once again in full analogy to the single-field case.

Finally, we turn our attention to the Einstein frame in generalized scalar-torsion theories, which we now define by the conditions
\begin{equation}
\ef{\mt{\mathcal{A}}} \equiv 1\,, \quad
\ef{\mathcal{Q}}_a \equiv 0\,,
\end{equation}
from which we deduce the transformation
\begin{equation}
\ef{\gamma} = \frac{1}{2}\ln\mt{\mathcal{A}}\,, \quad
\ef{\zeta}_a = \mathcal{Q}_a + \frac{\mt{\mathcal{A}}_{,a}}{2\mt{\mathcal{A}}}\,.
\end{equation}
Also here we see that the flatness constraint for the connection obstructs the existence of this frame unless \(\mathcal{Q}_{[a,b]} \equiv 0\). As in the previously studied multi-scalar-nonmetricity theories, we thus find the possibility to study a genuine extension to the pure multi-scalar-torsion class of theories. Only for the non-extended class we can define the invariant Einstein frame field variables as
\begin{equation}
\ef{g}_{\mu\nu} = \mt{\mathcal{A}}g_{\mu\nu}\,, \quad
\ef{\Gamma}^{\mu}{}_{\nu\rho} = \Gamma^{\mu}{}_{\nu\rho} + \left(\mathcal{Q}_a + \frac{\mt{\mathcal{A}}_{,a}}{2\mt{\mathcal{A}}}\right)\delta^{\mu}_{\nu}\phi^a_{,\rho}\,.
\end{equation}
Finally, the parameter functions in the action now become
\begin{equation}
\ef{\mt{\mathcal{B}}}_{ab} = 2\mt{\mathcal{F}}_{ab}\,, \quad
\ef{\mt{\mathcal{C}}}_a = 2\mt{\mathcal{H}}_a\,, \quad
\ef{\mt{\mathcal{V}}} = \mathcal{U}\,, \quad
\ef{\alpha} = \frac{1}{2}\ln\mathcal{I}\,, \quad
\ef{\beta}_a = \frac{\mathcal{I}_{,a}}{2\mathcal{I}} - \mathcal{P}_a\,.
\end{equation}
This concludes our discussion of frames in multi-scalar-teleparallel gravity theories. With these results at hand, we can now characterize a few subclasses of these theories independently of the choice of the frame. This will be done in the following section.

\section{Invariant characterization of theories}\label{sec:char}
In the previous sections we have studied the transformation of (multi-)scalar-teleparallel gravity theories under a particular class of field transformations and found a number of invariant quantities which describe these theories independently of the choice of their field variables. It is natural to expect that any physical predictions made by these theories are independent of the choice of these variables, and so we expect that these are fully determined in terms of invariants. In this section we will make use of these invariants in order to characterize a few subclasses of the general (multi-)scalar-teleparallel theories and discuss some of their properties. While a full phenomenological analysis of these theories would exceed the scope of this article, we aim to give at least a broad overview. The (multi-)scalar-teleparallel equivalent of (multi-)scalar-curvature gravity is discussed in section~\ref{ssec:stgequiv}. In section~\ref{ssec:boundcoup}, we include a coupling prescription of the scalar field to the teleparallel boundary term. In section~\ref{ssec:noder}, we discuss theories with no derivative coupling of the scalar field. Note that in all cases we keep the discussion general and consider an arbitrary number of scalar fields, while remarking on the single-field case if it has exhibits any significant difference from the multi-field case. It is therefore understood that all appearing parameter functions depend on the scalar field multiplet \(\boldsymbol{\phi}\).

\subsection{Scalar-teleparallel equivalent of scalar-curvature gravity}\label{ssec:stgequiv}
The different classes of (multi-)scalar-teleparallel theories we study in this article are motivated by a similar class of (multi-)scalar-curvature actions, which are defined by the action~\cite{Faraoni:2004pi,Fujii:2003pa}
\begin{equation}\label{eq:mscgaction}
S_{\text{SCG}} = \frac{1}{2\kappa^2}\int_M\left[\mathcal{A}\lc{R} + 2\mathcal{B}_{ab}X^{ab} - 2\kappa^2\mathcal{V}\right]\sqrt{-g}\dd^4x\,,
\end{equation}
where the only dynamical field variable is the metric. Hence, also the matter action has no coupling to the teleparallel connection, and so it takes the form
\begin{equation}
S_{\text{m}}\left[g_{\mu\nu}, \phi^a, \chi^I\right] = \hat{S}_{\text{m}}\left[g_{\mu\nu}e^{2\alpha}, \chi^I\right]\,.
\end{equation}
A crucial observation is the fact that the Ricci scalar \(\lc{R}\) of the Levi-Civita connection, which is the constituting ingredient of the Einstein-Hilbert action, is related to the teleparallel gravity scalar \(G\) (which reduces to \(Q\) in the symmetric teleparallel case and \(T\) in the metric teleparallel case) by a boundary term,
\begin{equation}
\lc{R} = -G + B\,, \quad
B = 2\lc{\nabla}_{\mu}M^{[\nu\mu]}{}_{\nu}\,,
\end{equation}
and so the (multi-)scalar-curvature action can be rewritten as
\begin{equation}\label{eq:mscgaction2}
S_{\text{SCG}} = \frac{1}{2\kappa^2}\int_M\left[\mathcal{A}(-G + B) + 2\mathcal{B}_{ab}X^{ab} - 2\kappa^2\mathcal{V}\right]\sqrt{-g}\dd^4x\,.
\end{equation}
Note that the term \(\mathcal{A}B\) is not a boundary term anymore, as it contains the function \(\mathcal{A}\) of the scalar fields. However, rewriting it using
\begin{equation}
\mathcal{A}\lc{\nabla}_{\mu}M^{[\nu\mu]}{}_{\nu} = \lc{\nabla}_{\mu}(\mathcal{A}M^{[\nu\mu]}{}_{\nu}) - \mathcal{A}_{,a}M^{[\nu\mu]}{}_{\nu}\lc{\nabla}_{\mu}\phi^a = \lc{\nabla}_{\mu}(\mathcal{A}M^{[\nu\mu]}{}_{\nu}) - \frac{1}{2}\mathcal{A}_{,a}(2U^a - V^a + W^a)\,,
\end{equation}
we see that up to a boundary term the action becomes equivalent to the action of the general teleparallel equivalent of scalar-curvature gravity, which reads~\cite{Hohmann:2022mlc}
\begin{equation}
S_{\text{GTESC}} = \frac{1}{2\kappa^2}\int_M\left[-\mathcal{A}G + 2\mathcal{B}_{ab}X^{ab} - \mathcal{A}_{,a}(2U^a - V^a + W^a) - 2\kappa^2\mathcal{V}\right]\sqrt{-g}\dd^4x\,.
\end{equation}
Comparing this action with the general multi-scalar-teleparallel action~\eqref{eq:mgenaction}, we see that it is a special case with the parameter functions given by
\begin{equation}
\mathcal{C}_a \equiv -2\mathcal{D}_a \equiv 2\mathcal{E}_a \equiv -\mathcal{A}_{,a}\,.
\end{equation}
These functions, of course, change if we perform a transformation of the metric and the teleparallel connection. However, we see by comparison with the definition~\eqref{eq:mvecinv2} that these conditions are equivalently written as
\begin{equation}
\mathcal{H}_a \equiv \mathcal{J}_a \equiv \mathcal{L}_a \equiv 0\,,
\end{equation}
and thus expressed in terms of invariants. We have thus found an invariant characterization of the equivalents of (multi-)scalar-curvature theories within the general class of (multi-)scalar-teleparallel theories of gravity.

It is straightforward to derive analogous conditions also for the generalized scalar-nonmetricity and scalar-torsion classes of gravity theories. If we include the Lagrange multiplier term~\eqref{eq:mtorlagmul} to constrain the torsion, we obtain the action
\begin{equation}
S_{\text{STESC}} = \frac{1}{2\kappa^2}\int_M\left[-\mathcal{A}Q + 2(\mathcal{B}_{ab} + 6\mathcal{A}_{,(a}\mathcal{T}_{b)} - 6\mathcal{A}\mathcal{T}_a\mathcal{T}_b)X^{ab} + (\mathcal{A}_{,a} - 2\mathcal{A}\mathcal{T}_a)(V^a - W^a) - 2\kappa^2\mathcal{V}\right]\sqrt{-g}\dd^4x
\end{equation}
of the generalized scalar-nonmetricity equivalent of scalar-curvature gravity, with the equivalent of scalar-curvature gravity given by the case \(\mathcal{T}_a \equiv 0\)~\cite{Jarv:2018bgs,Runkla:2018xrv}. In this case, we see that we find the subclass of multi-scalar-nonmetricity actions~\eqref{eq:msymaction} given by
\begin{equation}
2\st{\mathcal{D}}_a \equiv -2\st{\mathcal{E}}_a \equiv \st{\mathcal{A}}_{,a} - 2\st{\mathcal{A}}\mathcal{T}_a\,.
\end{equation}
We thus see that these theories are invariantly characterized by
\begin{equation}
\st{\mathcal{J}}_a \equiv \st{\mathcal{L}}_a \equiv 0\,.
\end{equation}
Finally, for the generalized scalar-torsion equivalent we include the Lagrange multiplier term~\eqref{eq:mnomlagmul} constraining the nonmetricity, which yields the action
\begin{equation}
S_{\text{MTESC}} = \frac{1}{2\kappa^2}\int_M\left[-\mathcal{A}T + 2(\mathcal{B}_{ab} - 6\mathcal{A}_{,(a}\mathcal{Q}_{b)} - 6\mathcal{A}\mathcal{Q}_a\mathcal{Q}_b)X^{ab} - 2(\mathcal{A}_{,a} + 2\mathcal{A}\mathcal{Q}_a)U^a - 2\kappa^2\mathcal{V}\right]\sqrt{-g}\dd^4x\,,
\end{equation}
with the scalar-torsion equivalent of scalar-curvature gravity given by \(\mathcal{Q}_a \equiv 0\)~\cite{Hohmann:2018ijr}. We thus see that a generalized multi-scalar-torsion theory belongs to this class if and only if its parameter functions satisfy the condition
\begin{equation}
-\mt{\mathcal{C}}_a \equiv \mt{\mathcal{A}}_{,a} + 2\mt{\mathcal{A}}\mathcal{Q}_a\,,
\end{equation}
which is expressed through invariant quantities as
\begin{equation}
\mt{\mathcal{H}}_a \equiv 0\,.
\end{equation}
This completes our discussion of scalar-curvature equivalents in all three families of scalar-teleparallel gravity theories.

\subsection{Theories with boundary term coupling}\label{ssec:boundcoup}
A straightforward generalization of the action~\eqref{eq:mscgaction2} of (multi-)scalar-curvature gravity is to introduce a different coupling function \(\mathcal{Z}\) for the boundary term, so that it takes the form
\begin{equation}\label{eq:mbcgaction}
S_{\text{BCG}} = \frac{1}{2\kappa^2}\int_M\left[-\mathcal{A}G + 2\mathcal{B}_{ab}X^{ab} - \mathcal{Z}B - 2\kappa^2\mathcal{V}\right]\sqrt{-g}\dd^4x\,.
\end{equation}
Once again performing integration by parts, as for the scalar-curvature equivalent discussed previously, one finds the corresponding general scalar-teleparallel equivalent action given by
\begin{equation}\label{eq:mgtebcaction}
S_{\text{GTEBC}} = \frac{1}{2\kappa^2}\int_M\left[-\mathcal{A}G + 2\mathcal{B}_{ab}X^{ab} + \mathcal{Z}_{,a}(2U^a - V^a + W^a) - 2\kappa^2\mathcal{V}\right]\sqrt{-g}\dd^4x\,.
\end{equation}
We see that only derivatives of \(\mathcal{Z}\) contribute to the action, and so adding any constant to this function does not change the action. This can also be seen from the original action~\eqref{eq:mbcgaction}, since any constant addition to \(\mathcal{Z}\) only contributes as a boundary term to the action. This ambiguity invites for another generalization of the action to become
\begin{equation}\label{eq:mgbcgaction}
S_{\text{GBCG}} = \frac{1}{2\kappa^2}\int_M\left[-\mathcal{A}G + 2\mathcal{B}_{ab}X^{ab} + \mathcal{Y}_{a}(2U^a - V^a + W^a) - 2\kappa^2\mathcal{V}\right]\sqrt{-g}\dd^4x\,,
\end{equation}
where the functions \(\mathcal{Y}_a\) are now arbitrary and not restricted to be derivatives of another function. This action was introduced in~\cite{Hohmann:2022mlc}, and its cosmology was studied in~\cite{Heisenberg:2022mbo}. By comparing this action with the general form~\eqref{eq:mgenaction}, we see that theories of this type are characterized by the conditions
\begin{equation}
\mathcal{C}_a \equiv -2\mathcal{D}_a \equiv 2\mathcal{E}_a\,.
\end{equation}
Comparing further with the definition~\eqref{eq:mvecinv2} of vector invariants, we see that this can be formulated invariantly as
\begin{equation}
\mathcal{H}_a \equiv -2\mathcal{J}_a \equiv 2\mathcal{L}_a\,.
\end{equation}
Similar results can also be obtained in the generalized (multi-)scalar-nonmetricity and (multi-)scalar-torsion classes of theories. Following the same procedure as discussed in detail in section~\ref{ssec:stgequiv}, which we do not repeat here for brevity, we find that a general multi-scalar-nonmetricity gravity theory has a generalized boundary term coupling defined by some arbitrary functions \(\mathcal{Y}_a\), and is thus of the form~\eqref{eq:mgbcgaction}, if and only if
\begin{equation}
\st{\mathcal{D}}_a + \st{\mathcal{E}}_a \equiv 0\,,
\end{equation}
which translates to the equivalent invariant condition
\begin{equation}
\st{\mathcal{J}}_a + \st{\mathcal{L}}_a \equiv 0\,.
\end{equation}
Finally, for the scalar-torsion case, we find that theories with generalized boundary term coupling already constitute the most general class of theories, since the only derivative coupling term \(U^a\) is obtained from a boundary term through integration by parts. We also remark that a theory is of the more restrictive form~\eqref{eq:mgtebcaction} if and only if the relevant vector invariants~\eqref{eq:mvecinv2} or~\eqref{eq:mvecinv4}, respectively, are integrable, i.e., given by the derivatives of some other (likewise invariant) function of the scalar fields. Note in particular that this is always true in the single-field case.

\subsection{Theories without derivative coupling}\label{ssec:noder}
It has been shown for all three flavors of scalar-teleparallel gravity theories that deviations of their post-Newtonian limit from GR, which become evident by parameters \(\beta\) and \(\gamma\) whose values deviate from their GR values \(\beta = \gamma = 1\), are introduced through the derivative coupling terms \(U^a, V^a, W^a\)~\cite{Emtsova:2019qsl,Flathmann:2019khc,Flathmann:2021itp,Hohmann:2023rqn}. Theories in which these terms do not appear in the gravitational action do not exhibit such deviations. In order to understand this condition, one must keep in mind that the post-Newtonian limit has been calculated in the Jordan frame, strictly imposing \(\alpha = 0\) and \(\beta_a = 0\). Hence, also the conditions on the parameter functions which characterize such minimally coupled theories and which are derived from the aforementioned post-Newtonian limit are formulated in the Jordan frame and read
\begin{equation}
\jf{\mathcal{C}}_a \equiv \jf{\mathcal{D}}_a \equiv \jf{\mathcal{E}}_a \equiv 0\,, \quad
\jf{\st{\mathcal{D}}}_a \equiv \jf{\st{\mathcal{E}}}_a \equiv 0\,, \quad
\jf{\mt{\mathcal{C}}}_a \equiv 0\,,
\end{equation}
depending on the particular class of theories under consideration. By comparison with the Jordan frame quantities listed in section~\ref{ssec:mframes}, we see that these correspond to the invariant conditions
\begin{equation}
\mathcal{K}_a \equiv \mathcal{M}_a \equiv \mathcal{N}_a \equiv 0\,, \quad
\st{\mathcal{M}}_a \equiv \st{\mathcal{N}}_a \equiv 0\,, \quad
\mt{\mathcal{K}}_a \equiv 0\,,
\end{equation}
respectively, for the three classes of theories. The invariance of the post-Newtonian limit under field transformations, which has been thoroughly studied in the (multi-)scalar-curvature case~\cite{Jarv:2014hma,Kuusk:2015dda}, becomes apparent by deriving the field equations, where one finds that contributions to the aforementioned post-Newtonian parameters arise from contributions to the scalar field equation. In the Jordan frame, these contributions are given only by derivatives of the teleparallel affine connection, while in other frames also the non-trivial matter coupling of the scalar fields contributes. It is the virtue of the invariant formalism that this contribution can be expressed independently of the choice of the field variables in terms of invariants only. Showing this explicitly throughout the field equations is a lengthy, but straightforward calculation; however, we will not enter this calculation here for brevity, and conclude with the remark that we expect also other observable properties of (multi-)scalar-teleparallel theories to be determined by invariant functions only.

\section{Conclusion}\label{sec:conclusion}
We have studied the behavior of different classes of (multi-)scalar-teleparallel gravity theories, which belong to the general, symmetric and metric types of teleparallel theories, under transformations of the three dynamical field variables: the metric, the connection and the scalar field(s). These transformations are defined by a number of functions of the scalar field(s). By comparing the action functionals for the original and the transformed field variables, we have identified three classes of theories, one of each of the aforementioned three types, which retain their form under these transformations. Any particular theory within these classes is defined by a number of functions of the scalar field(s), and field transformations relate theories which are defined by different choices of these free functions. Further, we have found several combinations of these functions which are invariant under the aforementioned transformations, and identified their role in the construction of analogues of the Jordan and Einstein frames known from scalar-curvature gravity theories. Finally, we have made use of the constructed invariants to characterize several subclasses of theories, which are either equivalent to already known theories, or have specific phenomenological properties.

One of our main results is the construction of generalized classes of scalar-torsion and scalar-nonmetricity gravity theories, in which nonmetricity or torsion are not imposed to vanish, but are algebraically constrained to be non-vanishing and determined by the scalar field in terms of new parameter functions \(\mathcal{Q}_a\) and \(\mathcal{T}_a\), respectively, in the Lagrange multiplier term. From the Bianchi identities, which show that vanishing curvature is compatible only with the conditions \(\mathcal{Q}_{[a,b]} \equiv 0\) or \(\mathcal{T}_{[a,b]} \equiv 0\), we found that these theories still maintain equivalence to the known ``pure'' scalar-torsion and scalar-nonmetricity theories. Theories which do not satisfy these conditions require also a non-vanishing, but algebraically constrained curvature and thus provide a genuine extension to the teleparallel framework.

The second main result is the construction of a set of quantities from the parameter functions defining the gravitational action which are invariant under transformations of the metric and the connection. By analogy with scalar-curvature gravity theories, in which a similar set of invariants has been constructed, we may conjecture that any physical properties of a given (multi-)scalar-teleparallel theory can be expressed solely in terms of these invariants. Further, by studying their behavior under transformations of the scalar field(s), in particular in the multi-field case, we have seen that they behave as tensors on the space of scalar fields. This finding supports the interpretation of the values of the scalar fields not simply as numbers, but as coordinates of a field space manifold, as it is known from so-called sigma models~\cite{Gell-Mann:1960mvl}. In fact, one of the invariants we constructed can indeed be interpreted as a Riemannian metric on the scalar field space, which is one of the typical ingredients of a sigma model.

Our results invite for numerous further studies. Given a genuine extension of scalar-torsion and scalar-nonmetricity theories of gravity, the question for their phenomenological properties obviously suggests itself. In particular, one may pose the question whether these are able to address open questions such as those arising from cosmology or the strong coupling problem, while still maintaining consistency with the post-Newtonian limit in the solar system. The invariant quantities we introduced in this work may become a useful tool in such future studies, and it has to be investigated whether the aforementioned conjecture, that the physical properties of a given theory only depend on these invariants, holds for observables at the classical level as well as for a possible quantum extension. Another intriguing possibility for future investigations stems from the interpretation of the scalar field space as a manifold, thus allowing for a non-trivial topology, and in particular non-vanishing first de Rham cohomology \(H_{\text{dR}}^1\). In this case one may have parameter functions \(\mathcal{Q}_a\) and \(\mathcal{T}_a\) which satisfy \(\mathcal{Q}_{[a,b]} \equiv 0\) or \(\mathcal{T}_{[a,b]} \equiv 0\), hence they constitute the components of closed one-forms, but are not exact. Another possibility, which may likewise benefit from the formalism of invariants and their geometric interpretation, is to study ``teleparallel'' gravity theories with non-vanishing, but algebraically constrained curvature, as well as their ability to address the aforementioned open questions. Finally, one may also aim to generalize these results by considering the Horndeski class of scalar-teleparallel theories~\cite{Bahamonde:2019shr,Bahamonde:2019ipm,Bahamonde:2020cfv,Bahamonde:2021dqn,Bahamonde:2022cmz}, which is motivated by the Horndeski class of scalar-curvature gravity theories~\cite{Horndeski:1974wa,Deffayet:2011gz,Kobayashi:2011nu,Gleyzes:2014dya,Kobayashi:2019hrl}, and extend the studied field transformations to include also disformal transformations~\cite{Bekenstein:1992pj,Bettoni:2013diz,Ezquiaga:2017ner,Zumalacarregui:2013pma,Hohmann:2019gmt,Golovnev:2019kcf}.

\begin{acknowledgments}
MH gratefully acknowledges the full financial support by the Estonian Research Council through the Personal Research Funding project PRG356.
\end{acknowledgments}

\bibliography{scalarteletrans}
\end{document}